\documentclass[lettersize,journal]{IEEEtran}
\usepackage{amsmath,amsfonts}
\usepackage{algorithmic}
\usepackage{array}
\usepackage[caption=false,font=normalsize,labelfont=sf,textfont=sf]{subfig}
\usepackage{textcomp}
\usepackage{stfloats}
\usepackage{url}
\usepackage{verbatim}
\usepackage{graphicx}
\usepackage[colorlinks,linkcolor=blue,anchorcolor=blue,citecolor=blue,urlcolor=blue]{hyperref}
\def\BibTeX{{\rm B\kern-.05em{\sc i\kern-.025em b}\kern-.08em
    T\kern-.1667em\lower.7ex\hbox{E}\kern-.125emX}}
\usepackage{balance}
\usepackage{cite}
\usepackage{orcidlink}
\usepackage{wrapfig}

\usepackage{bbm}
\usepackage{soul,xcolor}

\usepackage[figuresright]{rotating}
\usepackage{amssymb}
\usepackage{amsmath}
\usepackage{amsthm}
\usepackage[shortlabels]{enumitem}
\usepackage{mathtools}
\usepackage{dcolumn}
\usepackage{physics}
\usepackage{float}
\usepackage{bm}
\usepackage{braket}
\usepackage[normalem]{ulem}
\usepackage[ruled,vlined,linesnumbered]{algorithm2e}
\usepackage{thm-restate}
\usepackage{setspace}
\usepackage{stackengine}
\usepackage{qcircuit}

\usepackage{multirow}
\usepackage{diagbox}

\usepackage{cuted}

\usepackage{fancyhdr}

\usepackage{xr-hyper}
\usepackage[colorlinks,linkcolor=blue,anchorcolor=blue,citecolor=blue,urlcolor=blue]{hyperref}
\makeatletter

\newcommand*{\addFileDependency}[1]{
\typeout{(#1)}
\@addtofilelist{#1}
\IfFileExists{#1}{}{\typeout{No file #1.}}
}\makeatother

\newcommand*{\myexternaldocument}[1]{%
\externaldocument{#1}%
\addFileDependency{#1.tex}%
\addFileDependency{#1.aux}%
}
\myexternaldocument{supplement}

\newcommand{\JM}[1]{{\textcolor{cyan}{John: #1}}}
\newcommand{\kcs}[1]{{\textcolor{purple}{Kevin: #1}}}

\newcommand{\papertitle}{Toward Mixed Analog-Digital Quantum Signal Processing: Quantum AD/DA Conversion and the Fourier Transform}

\begin{document}

\title{\papertitle}

\author{
Yuan Liu*\orcidlink{0000-0003-1468-942X}, \IEEEmembership{Member, IEEE},
John M.~Martyn*\orcidlink{0000-0002-4065-6974}, Jasmine Sinanan-Singh, Kevin C. Smith\orcidlink{0000-0002-2397-1518},
Steven M. Girvin\orcidlink{0000-0002-6470-5494}, and
Isaac L.~Chuang\orcidlink{0000-0001-7296-523X}
\thanks{Manuscript created December, 2023. *These authors contributed equally. (\textit{Corresponding author: Yuan Liu.})}
\thanks{Y.~Liu is with the Departments of Electrical and Computer Engineering, Computer Science, and Physics (affiliated), North Carolina State University, Raleigh, North Carolina 27606, USA. He was also with the Department of Physics, Research Laboratory of Electronics, Massachusetts Institute of Technology, Cambridge, Massachusetts 02139 USA (e-mail: q\_yuanliu@ncsu.edu).}
\thanks{J.M.~Martyn is with the Department of Physics, Massachusetts Institute of Technology, Cambridge, Massachusetts 02139, USA (e-mail: jmmartyn@mit.edu).}
\thanks{J.~Sinanan-Singh is with the Department of Physics, Massachusetts Institute of Technology, Cambridge, Massachusetts 02139, USA (e-mail: jsinanan@mit.edu).}
\thanks{K.C.~Smith was with the Department of Physics and the Yale Quantum Institute, Yale University, New Haven, Connecticut 06511 USA and Brookhaven National Laboratory, Upton, New York 11973, USA (e-mail: kevin.smith@yale.edu).}
\thanks{S.M.~Girvin is with the Department of Physics and the Yale Quantum Institute, Yale University, New Haven, Connecticut 06511, USA (e-mail: steven.girvin@yale.edu).}
\thanks{I.L.~Chuang is with the the Department of Physics, and the Department of Electrical Engineering and Computer Science, Massachusetts Institute of Technology, Cambridge, Massachusetts 02139, USA (e-mail: ichuang@mit.edu).}
}


\maketitle

\begin{abstract}
Signal processing stands as a pillar of classical computation and modern information technology, applicable to both analog and digital signals. Recently, advancements in quantum information science have suggested that quantum signal processing (QSP) can enable more powerful signal processing capabilities. However, the developments in QSP have primarily leveraged \emph{digital} quantum resources, such as discrete-variable (DV) systems like qubits, rather than \emph{analog} quantum resources, such as continuous-variable (CV) systems like quantum oscillators. Consequently, there remains a gap in understanding how signal processing can be performed on hybrid CV-DV quantum computers. Here we address this gap by developing a new paradigm of mixed analog-digital QSP. We demonstrate the utility of this paradigm by showcasing how it naturally enables analog-digital conversion of quantum signals--- specifically, the transfer of states between DV and CV quantum systems. We then show that such quantum analog-digital conversion enables new implementations of quantum algorithms on CV-DV hardware. This is exemplified by realizing the quantum Fourier transform of a state encoded on qubits via the free-evolution of a quantum oscillator, albeit with a runtime exponential in the number of qubits due to information theoretic arguments. Collectively, this work marks a significant step forward in hybrid CV-DV quantum computation, providing a foundation for scalable analog-digital signal processing on quantum processors. 
\end{abstract}


\section{Introduction}

The ability to process signals in an efficient and robust manner is a cornerstone of engineering and technology, from audio and speech recognition, to computer design and communications~\cite{proakis2007digital,oppenheim1999discrete}. In the wake of modern computers, sophisticated frameworks and algorithms have been developed to process classical signals, including the fast Fourier transform~\cite{van1992computational}, Shannon sampling~\cite{marks2012introduction}, and filter design~\cite{winder2002analog}. 

Classical signal processing has benefited from both \emph{digital} and \emph{analog} computing devices~\cite{haensch2018next}. While digital signal processing enables the processing of discretized signals, analog signal processing is used for processing continuous signals, such as audio and speech data~\cite{gold2011speech}. Hybrid analog-digital computing~\cite{kester2003mixed} has also shown great promise, with notable applications in improving energy efficiency~\cite{sharpeshkar2010ultra, guo2016energy}.

In contrast to the classical setting, quantum systems, governed by the laws of quantum mechanics, exhibit fundamentally different behavior than their classical counterparts and have been shown to facilitate more powerful models of computation than classical computers~\cite{Nielsen_Chuang}. It is therefore natural to ask if quantum computation can process signals more efficiently and powerfully than classical methods.\footnote{In posing this question, we are interested in processing quantum signals (quantum amplitudes) using quantum resources, rather than processing classical signals on a quantum computer.} 

A pioneering work in this direction~\cite{eldar2002quantum} introduced a quantum-inspired signal processing paradigm based on quantum-mechanical concepts, to design an array of novel classical signal processing methods. Extending this line of research, Refs.~\cite{low2016methodology,low2017optimal} proposed \emph{quantum signal processing} (QSP) as a quantum algorithm that enables the design and implementation of a polynomial transformation of a quantum amplitude. QSP has since been generalized to transform a linear operator embedded in a larger Hilbert space, leading to the celebrated quantum singular value transformation (QSVT)~\cite{gilyen2019quantum}. As an illustration of the power of QSP and QSVT, Ref.~\cite{GrandUnificationAlgos} shows how major quantum algorithms, including Grover search, Shor's factoring algorithm, and Hamiltonian simulation, can all be realized through QSVT. Inspired by this remarkable progress, a number of recent works have further generalized QSP and QSVT~\cite{rossi2021mqsp,motlagh2023generalized,rossi2023quantum}, studied the noise robustness of QSP~\cite{tan2023perturbative}, presented algorithms for efficient computation of QSP/QSVT transformations~\cite{chao2020finding,WhaleyRobust,ying2022stable}, and showcased applications of QSP/QSVT to relevant problems~\cite{martyn2023efficient,dong_beyond_2022}. Moreover, by combining quantum computing techniques and classical signal processing, recent work on quantum-enhanced signal processing \cite{10887919,10434630} also demonstrated the potential of quantum computing in the broader areas of signal processing and communication \cite{10721632,10593013,10670345,shukla2022hybrid}.

These developments in QSP rely on the capabilities of digital quantum computers, in which quantum states are encoded in discrete-variable (DV) systems, i.e., qubits (or possibly qudits). Separate from these DV systems, continuous-variable (CV) quantum systems, such as the quantum harmonic oscillator, are ubiquitous in practice and also provide useful quantum resources \cite{liu2016power,Braunstein2005}. A prominent example of such CV systems is the electromagnetic (EM) wave used in wireless and communications \cite{mikki2020quantum}, which obey classical wave mechanics at high intensity, but exhibit quantum effects at low intensity. Recent experimental progress in the control and engineering of CV quantum systems has made them essential to quantum information science, prompting efforts to harness CV systems for computation~\cite{Campagne-Ibarcq2020,Sivak_GKP_2022,LuyanSun2020,ni2022beating,EickbuschECD,Braunstein2005,Weedbrook2012,andersen2015hybrid,Krastanov2015,QuditsfromOscillatorsPhysRevA.104.032605}. In this direction, recent works have developed hybrid CV-DV quantum processors~\cite{andersen2015hybrid,liu2024hybrid}, which combine DV and CV quantum systems into a powerful new framework for quantum computation, with natural applications to problems such as simulating coupled fermion-boson systems \cite{C2QA-LGTpaper}.

Because QSP algorithms have primarily leveraged DV quantum systems, we currently lack the ability to use CV systems and hybrid CV-DV processors in this context. This is in stark contrast to classical signal processing, which has profited from both analog and digital modes of computation. 
A major challenge in extending QSP to CV and CV-DV systems is the drastic differences between DV and CV quantum states. For instance, while DV quantum states have finite dimensionality, CV quantum states have infinite dimensionality and are supported over the entire real axis in position space.

In this work, we address this challenge by establishing a framework of \emph{mixed analog-digital QSP}, which enables processing of quantum signals on hybrid CV-DV quantum hardware. Our framework encompasses two signal processing primitives: (1) hybrid single-variable QSP for constructing polynomial transformations of either position $\hat{x}$ \emph{or} momentum $\hat{p}$, and (2) hybrid non-Abelian QSP for constructing polynomial transformations of both $\hat{x}$ \emph{and} $\hat{p}$. We use ``non-Abelian" to refer to the fact that the quantum mechanical operators $\hat{x}$ and $\hat{p}$ do not commute, i.e., $[\hat{x}, \hat{p}] = \hat{x} \hat{p} - \hat{p}\hat{x} \ne 0$.
The polynomials achievable with the hybrid single-variable QSP are characterized by the QSP theorems established in Refs.~\cite{qspi2023,low2016methodology}, whereas the polynomials achievable with hybrid non-Abelian QSP are in principle more powerful~\cite{liu2024hybrid, singh2025nonabelian} but a complete theory has yet to be established.

The ability to perform QSP on CV-DV hardware provides a cookbook for implementing quantum algorithms on hybrid quantum processors. To facilitate the development of these algorithms, one requires a mechanism to reliably convert between CV states and DV states. It is known that such analog-to-digital (AD) and digital-to-analog (DA) conversion can be accomplished in classical signal processing via sampling and interpolation, respectively \cite{marvasti2012unified,4472240,grochenig2001foundations,1003065}. However, a quantum analogue of these concepts is not obvious~\cite{kitaev2008wavefunction}.
A key distinction between classical and quantum AD/DA conversion is that the quantum case must be unentangling: the input and output states must be unentangled across the DV and CV systems to ensure full transmission of information from one system to the other. Any remaining entanglement in the final state implies that information in the initial state will remain in the quantum correlations between the systems, inaccessible to either party individually. As the concept of entanglement does not exist classically, this requirement must be treated with care. 
Here we show how quantum AD/DA conversion is naturally enabled by mixed analog-digital QSP. In particular, we illustrate two protocols that transfer a DV state into an equivalent CV state, and vice-versa, and provide analytical error bounds on their performance. The first protocol is constructed with hybrid single-variable QSP, while the second protocol, initially introduced in Ref.~\cite{hastrup2021improved}, can be recontextualized as an instance of non-Abelian hybrid QSP. We note the extensive literature on advanced sampling techniques such as non-uniform sampling, compressive sensing, and sparse signal processing in the classical domain \cite{aldroubi2001nonuniform,marvasti2012unified,4472240,grochenig2001foundations,1003065,baron2009bayesian}, we focus this first work in the quantum realm on the uniform sampling case, and leave the non-uniform case for future work.

As an application of this paradigm, we use quantum AD/DA conversion protocols to implement the quantum Fourier transform of a state on qubits by using the natural dynamics (free-evolution) of an oscillator. This is realized in three steps: 1) transferring the initial state on qubits to an oscillator; 2) performing free-evolution of the oscillator; 3) and transferring the oscillator state back to qubits. Importantly, our construction is fully coherent and does not require post-selection, in contrast to an alternative construction put forth in Ref.~\cite{chen2019quantum}. From a signal processing perspective due to the Nyquist criteria, the required CV oscillator phase space area (analog signals) must be proportional to the Hilbert space dimension of the DV system (number of data points for digital signals) to avoid significant loss of information for general quantum states. This renders the runtime of our protocol (or physical resources required) necessarily linear in the dimension of the DV Hilbert space (i.e. exponential in the number of qubits), despite the fact that the gate count can still scale polynomially. We emphasize that our work is entirely different from existing work on performing the quantum Fourier transform in the so-called digital-analog quantum computation model which combines digital quantum computation with analog quantum simulators \cite{PhysRevResearch.2.013012}.

The rest of the paper is organized as follows. We first review basics of signal processing and hybrid DV-CV quantum systems in Sec.~\ref{sec:qft-osc-review}, and subsequently present mixed analog-digital QSP in Sec.~\ref{sec:hybrid-qsp}, including hybrid single-variable QSP and non-Abelian hybrid QSP. Then in Sec.~\ref{sec:state-transfer-sampling}, we present two protocols for quantum AD/DA conversion. Armed with hybrid QSP and quantum AD/DA conversion, in Sec.~\ref{sec:qft} we show how to realize the Fourier transform of a DV quantum state by the free evolution of a CV system. Finally, in Sec.~\ref{sec:conclusion-outlook} we conclude and discuss the outlook of this work.

\textbf{Notation:} In this paper, we will denote classical vectors with boldface as $\mathbf{v}$, and quantum states with kets as $|\psi \rangle$. We will use a subscript $Q$ (i.e., $|\psi \rangle_Q$) to denote a state on qubits, and a subscript $O$ (i.e., $|\psi\rangle_O$) to denote a state on an oscillator. Lastly, we will denote quantum operations on the CV system with hats, e.g., $\hat{x}$ for position.

\section{Overview of Signal Processing and CV-DV Quantum Systems}
\label{sec:qft-osc-review}
\noindent 

In this section, we present the concepts that underlie classical signal processing and quantum signal processing, with the aim of familiarizing readers from either community with the necessary background to understand this work. Sec.~\ref{ssec:overview-classical-sp} overviews classical signal processing, and highlights the continuous-discrete and periodic-aperiodic nature of classical signals. Thereafter, Sec.~\ref{ssec:basic-hybrid-dvcv} introduces the basics of DV and CV quantum systems and operations.

\subsection{Overview of classical signal processing data types}
\label{ssec:overview-classical-sp}

\begin{figure}
    \centering
    \includegraphics[width=0.48\textwidth]{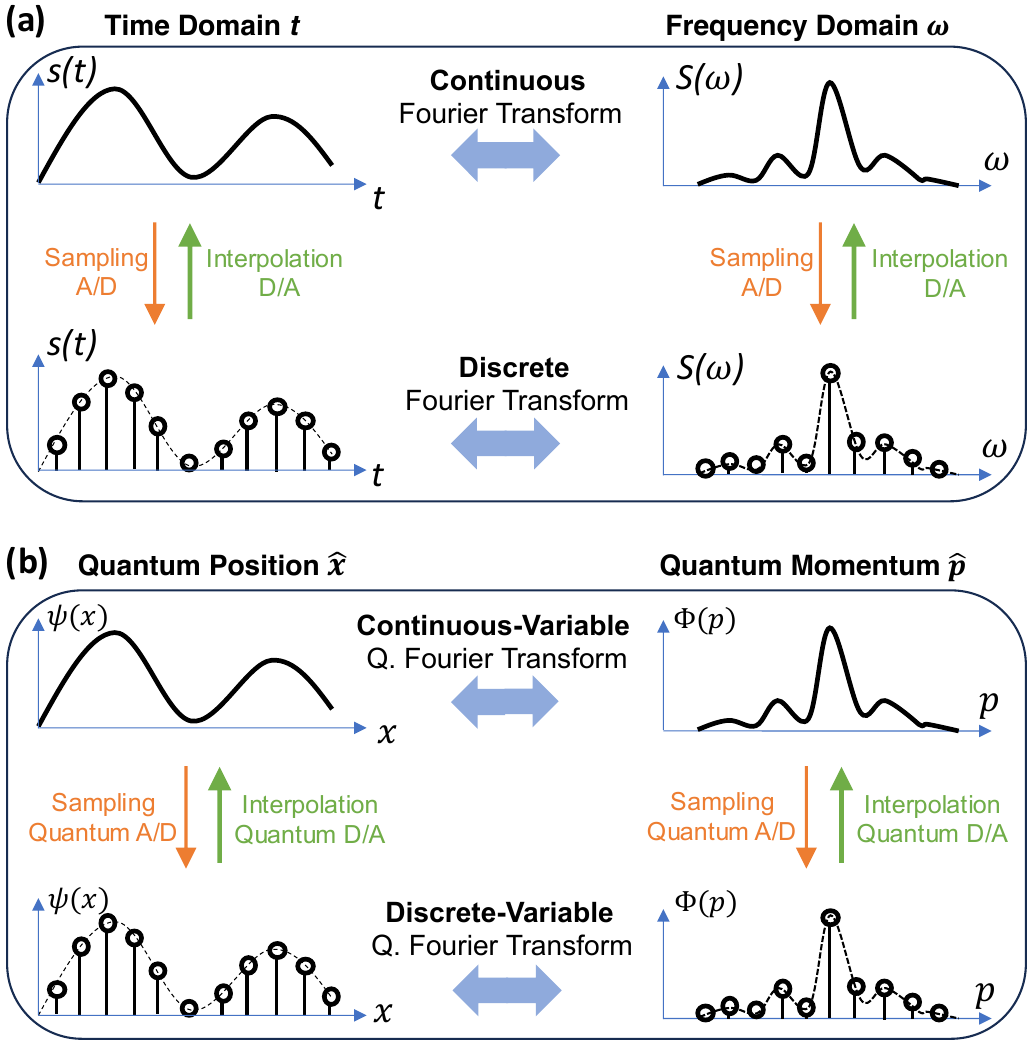}
    \caption{Schematic of the duality between \textbf{(a)} time-frequency-domain classical analog-digital signals versus \textbf{(b)} quantum position and momentum domain CV-DV signals. This work develops quantum AD/DA conversion protocols in panel (b) to facilitate mixed analog-digital \emph{quantum} signal processing, in similar spirit to classical sampling/interpolation for mixed analog-digital \emph{classical} signal processing in panel (a).}
    \label{fig:CV-DV-schematic}
\end{figure}

We begin by summarizing classical signal types in Fig.~\ref{fig:CV-DV-schematic}(a), including the relationships between them. In classical systems, a \emph{signal} is a physical quantity that is a real-valued continuous function of time, such as electric current or voltage, although complex representations can be used to ease mathematical description. These analog signals can be converted to the frequency domain by the continuous Fourier transform (upper panel of Fig.~\ref{fig:CV-DV-schematic}(a)), and processed with analog filters. 
To improve the robustness of signal processing, continuous signals are often ``quantized"\footnote{Despite the name, this has no relation to quantum mechanics.} into discrete signals via \emph{sampling}, or equivalently analog-digital (A/D) conversion. In this context, a discrete signal may be transformed to frequency domain by the discrete Fourier transform (lower panel of Fig.~\ref{fig:CV-DV-schematic}(1)). Inversely, discrete signals can also be converted into continuous signals via \emph{interpolation}, or digital-analog (D/A) conversion. In addition to being continuous or discrete, signals can also be \emph{periodic} or \emph{aperiodic}, and often require correspondingly different treatments. 
To accommodate these differences, windowing and padding techniques are used to connect signal processing tasks between signals of different periodicity. 

Depending on whether the time-domain signal is periodic or aperiodic, and continuous or discrete, there are four possible signal types. Similarly, the corresponding frequency domain signals also come in four types. As a result, there are sixteen possible transformations that connect time-domain signals to their frequency domain counterparts. 
These transformations are named according to the characteristics of the signals they transform between. For example, the continuous Fourier transform connects a continuous aperiodic signal in the time domain to a frequency-domain signal that is also continuous and aperiodic. Likewise, padding/windowing techniques and sampling/interpolation transform a continuous aperiodic signal in the time domain to a discrete periodic signal in the frequency domain. 
Other transformations can be analogously defined to connect any pair of time- and frequency-domain signals; see Refs.~\cite{proakis2007digital,oppenheim1999discrete} for more details.

Beyond processing analog and digital signals on classical computers, recent developments in quantum computing and engineering raise the question: can mixed analog-digital signal processing be achieved on \emph{quantum systems}? And if so, how can one define notions of \emph{quantum sampling and interpolation} to bridge analog and digital \emph{quantum} data, and how can we use these methods to implement algorithms like the Fourier transform? Addressing these questions is crucial but challenging due to the fundamental differences between quantum and classical systems.
As we will see later in this paper, we obtain affirmative answers to these questions by utilizing the fact that time-frequency duality is mathematically similar to position-momentum duality in quantum mechanics (Fig. \ref{fig:CV-DV-schematic}(b)). Understanding our results and resolution requires a strong background on discrete and continuous variable quantum systems, to which we now turn.

\subsection{Review of DV and CV Quantum Systems and Operations}
\label{ssec:basic-hybrid-dvcv}
Here we review the basics of discrete-variable (Sec.~\ref{sec:gates-operations-qubit}) and continuous-variable (Sec.~\ref{sec:gates-operations-osc}) quantum systems, including the quantum gates operations achievable on these systems.

\subsubsection{Quantum States and Operations on Qubits}
\label{sec:gates-operations-qubit}

A qubit is a two-level quantum system whose state can be represented as a linear combination of two orthonormal basis states, denoted by $\ket{0}$ and $\ket{1}$. That is, an arbitrary state can be written as  $\ket{\psi} = c_0 \ket{0} + c_1 \ket{1}$, for $c_0, c_1 \in \mathbb{C}$, where it is normalized as $|c_0|^2 + |c_1|^2 = 1$. Conventionally, we write $\ket{0} = [1,0]^T$, $\ket{1} = [0,1]^T$, and $\ket{\psi} = [c_0, c_1]^T$. 

To transform a one qubit state $\ket{\psi}$ to another $\ket{\psi'}$, we require a single-qubit gate, denoted $R_{\bm{\hat{b}}}(\theta)$, such that $\ket{\psi'} = R_{\bm{\hat{b}}}(\theta) \ket{\psi}$. In general, $R_{\bm{\hat{b}}}(\theta)$ can be an arbitrary $2\times 2$ special unitary matrix (SU(2)), 
\begin{align}
    R_{\bm{\hat{b}}}(\theta) = e^{- i \frac{\theta}{2} \bm{\hat{b}} \cdot \bm{\sigma}},
    \label{eq:su2}
\end{align}
which is generated by a set of $2 \times 2$ Hermitian matrices known as the Pauli matrices: $
    \sigma_x = 
    \begin{bmatrix}
        0 & 1 \\
        1 & 0
    \end{bmatrix}, ~~ 
    \sigma_y = 
    \begin{bmatrix}
        0 & -i \\
        i & 0
    \end{bmatrix}, ~~
    \sigma_z = 
    \begin{bmatrix}
        1 & 0 \\
        0 & -1
    \end{bmatrix}
$
where $i = \sqrt{-1}$, and $\bm{\sigma} = [\sigma_x, \sigma_y, \sigma_z]$ is a vector of the Pauli matrices, $\theta \in [0, 4 \pi)$ is a rotation angle, and $\bm{\hat{b}} = [b_x, b_y, b_z] \in \mathbb{R}^3$ is a vector of unit length, $|b_x|^2 + |b_y|^2 + |b_z|^2 = 1$. In general, $\theta$ and $\bm{\hat{b}}$ can be chosen from a continuum of possibilities to realize an arbitrary single-qubit gate. 

To apply quantum computation to multiple qubits, additional gates that \emph{entangle} qubits are needed. One common such gate is the controlled-NOT (CNOT) operation, which acts on two qubits as ${\rm CNOT} = \frac{I + \sigma_z}{2} \otimes I + \frac{I - \sigma_z}{2} \otimes \sigma_x$ where $I$ is the $2\times 2$ identity matrix, and $\otimes$ is the tensor product. 
It can be shown~\cite{Nielsen_Chuang} that the gate set $\mathcal{S}_0 = \{R_{\bm{\hat{b}}}(\theta), {\rm CNOT} \}$ forms a \emph{universal} gate set, such that an arbitrary gate on $n$ qubits (i.e., a $2^n \times 2^n$ unitary matrix) can be decomposed into a product of $R_{\bm{\hat{b}}}(\theta)$ and ${\rm CNOT}$ gates.

Despite its universality, the set $\mathcal{S}_0$ contains infinitely many gates due to the continuous parameterization of $R_{\bm{\hat{b}}}(\theta)$. It turns out an arbitrary gate $R_{\bm{\hat{b}}}(\theta)$ can be decomposed into a finite sequence of gates from the discrete set $\{H, T\}$, where
\begin{equation}
    H = \frac{1}{\sqrt{2}} \begin{bmatrix} 1 & 1 \\ 1 & -1 \end{bmatrix}, \quad T = \begin{bmatrix} 1 & 0 \\ 0 & e^{i \frac{\pi}{4}} \end{bmatrix}
\end{equation}
are the Hadamard gate and $T$ gate, respectively. This implies that the set $\mathcal{S}_1 = \{H, T, {\rm CNOT} \}$ is a universal gate set for qubit-based quantum computation~\cite{boykin2000new}. According to the Solovay-Kitaev theorem, constructing an $\epsilon$-approximation to arbitrary gate $R_{\bm{\hat{b}}}(\theta)$ requires $O(\log^c(\frac{1}{\epsilon}))$ $H$ or $T$ gates, where $c$ is a constant close to $2$~\cite{dawson2005solovay}. As a byproduct of this result, synthesizing an arbitrary $n$-qubit unitary requires $O(4^n \log^c(\frac{1}{\epsilon}))$ gates from the set $\mathcal{S}_1$~\cite{dawson2005solovay}.

\subsubsection{Quantum States and Operations on Oscillators}
\label{sec:gates-operations-osc}
In contrast to qubits, the computational capabilities of continuous variable quantum systems are less well-studied (for an introduction see \cite{liu2024hybrid}). Here we will consider the paradigmatic continuous variable system --- the quantum harmonic oscillator. The quantum harmonic oscillator arises in any quantum system that exhibits oscillations, such as molecular vibrations, microwave photons in cavities, and phonons in solids \cite{liu2024hybrid}. Its Hamiltonian is given by the quantized version of the familiar classical harmonic oscillator
\begin{equation}
    H_0 = \frac{p^2}{2 m} + \frac{1}{2} m \omega_0^2 x^2 ,
    \label{eq:h0-def}
\end{equation}
where $p$ and $x$ are the momentum and position, and $m$ and $\omega_0$ are the mass and frequency of the oscillator. 
Mathematically, the quantization of the harmonic oscillator is achieved by introducing momentum and position operators $\hat{p}$ and $\hat{x}$ that obey the canonical commutation relation $[\hat{x}, \hat{p}] := \hat{x} \hat{p} - \hat{p} \hat{x} = i$.
Setting $\hbar=1$ and $m\omega_0 = 1$ for simplicity, these operators can be conveniently expressed in terms of the \emph{annihilation and creation operators}, $a, a^\dagger$, respectively, defined as
\begin{equation}
    \hat{p} = \frac{i(a^\dagger - a)}{\sqrt{2}}\, , %
~~~~~~~~~
    \hat{x} = \frac{a + a^\dagger}{\sqrt{2}} \label{eq:x-def}, 
\end{equation}
and obeying $[a, a^\dagger] = 1$.

After quantization, the Hamiltonian of the quantum harmonic oscillator is given by $\hat{H}_0 = \omega_0 (\hat{n} + \frac{1}{2})$ where $\hat{n} := a^\dagger a$ is known as the number operator. The number operator has non-negative integer eigenvalues $n\in \mathbb{Z}_0^+$, which correspond to number of excitations in the oscillator. Incidentally, the corresponding eigenstates, which we denote by $|n\rangle$, are also the eigenstates of the Hamiltonian
\begin{align}
    \hat{H}_0 \ket{n} = E_n \ket{n},
\end{align}
with eigenvalues (i.e., energy) $E_n = \omega_0 (n+\frac{1}{2})$ for $n = 0, 1, 2, \ldots $. Similar to the qubit case, an arbitrary oscillator state can be written as a linear combination of $\ket{n}$ as $\ket{\psi} = \sum_{n=0}^\infty c_n \ket{n}$ for $c_n \in \mathbb{C}$ and $\sum_{n=0}^\infty |c_n|^2 = 1$. Adjacent eigenstates are related to each other by $a, a^\dagger$ as:
\begin{align} \label{eq:adagger-def}
    a^\dagger \ket{n} = \sqrt{n+1} \ket{n+1}, \qquad  
    a \ket{n} = \sqrt{n} \ket{n-1}. 
\end{align}
In the language of linear algebra, $|n\rangle$ can be represented by a (infinite-dimensional) column vector with its $n$-th entry set to 1 and the rest to 0; for example, $\ket{0} = \begin{bmatrix}
    1 & 0 & 0 & \cdots
\end{bmatrix}^T$, $\ket{1} = \begin{bmatrix}
    0 & 1 & 0 & \cdots
\end{bmatrix}^T$, and $\ket{2} = \begin{bmatrix}
    0 & 0 & 1 & 0 & \cdots
\end{bmatrix}^T$. In this basis, Eqs.~\eqref{eq:adagger-def} lend themselves to the following matrix representation for $\hat{p}, \hat{x}$:
\begin{align}
    \hat{p} &= \frac{i}{\sqrt{2}}
    \begin{bmatrix}
        0&-1&0&0& \cdots \\
        1&0&-\sqrt{2}&0& \cdots \\
        0&\sqrt{2}&0&-\sqrt{3}&\cdots\\
        0&0&\sqrt{3}&0&\cdots\\
        \vdots & \vdots &\vdots& \vdots &\ddots
    \end{bmatrix} , \quad 
    \\
    \hat{x} &= \frac{1}{\sqrt{2}}
    \begin{bmatrix}
        0&1&0&0& \cdots \\
        1&0&\sqrt{2}&0& \cdots \\
        0&\sqrt{2}&0&\sqrt{3}&\cdots\\
        0&0&\sqrt{3}&0& \cdots \\
        \vdots & \vdots &\vdots& \vdots &\ddots
    \end{bmatrix} .
\end{align}

Just as the Pauli matrices generate an arbitrary single-qubit operation as per Eq.~\eqref{eq:su2}, an arbitrary unitary operation on an oscillator can be represented as
\begin{align}
    U = e^{-i h(\hat{x}, \hat{p})},
    \label{eq:osc-gate-def}
\end{align}
where $h(\hat{x}, \hat{p})$ is a function of $\hat{x}, \hat{p}$, often expressed as a polynomial in $\hat{x}$ and $\hat{p}$. Universal control of an oscillator requires the ability to implement the above unitary for any such $h(\hat{x}, \hat{p})$ of finite degree. In a similar manner, an entangling gate between two oscillators can be defined to achieve universal quantum computation on multiple oscillators. See Ref. \cite{liu2024hybrid} for more details on universal gate sets (continuously parametrized) for oscillator and hybrid oscillator-qubit quantum computation.

Despite the parallel discussion of Eq.~\eqref{eq:su2} and Eq.~\eqref{eq:osc-gate-def}, qubit gates and oscillator gates differ significantly. For example, in the qubit case, any even power of a Pauli matrix is the identity (e.g., $\sigma_x^2 = 1$), whereas in the oscillator case powers of $\hat{x}, \hat{p}$ are non-trivial. This distinction arises from the infinite dimensionality of the oscillator, and complicates the construction of unitary operations on CV systems. 

Nonetheless, a simple example of such a unitary operator on a CV system is the free-evolution of an oscillator, given by
\begin{align}\label{eq:Fourier_gate}
    R(\theta) = e^{- i \theta a^\dagger a},
\end{align}
where $\theta = \omega_0 t$ is a rotation angle. When $\theta = \frac{\pi}{2}$, this becomes the ``Fourier gate" $F = R(\frac{\pi}{2})$, which acts as 
\begin{align} \label{eq:Fourier_gate2}
    F^\dagger \hat{x} F &= \hat{p}, \qquad 
    F^\dagger \hat{p} F = -\hat{x}.
\end{align}
Evidently, the Fourier gate swaps the position $\hat{x}$ with the momentum operator $\hat{p}$, thus enacting a continuous Fourier transform on the underlying wave function (Fig. \ref{fig:CV-DV-schematic}(b)).

Lastly, we note that in practical/numerical studies of CV systems, the infinite-dimensional Hilbert space can be truncated to a finite dimension by selecting a large integer $N_{\rm max}$ to represent the maximum excitation number. While this strategy can in principle introduce significant error for arbitrary CV-DV quantum circuits, physically implementable operations produce states with finite energy; consequently, such circuits (including those considered here) can be well-approximated through an appropriate, problem-specific choice of $N_{\rm max}$ \cite{liu2024hybrid}. Truncating the states and operations past this cutoff renders the system finite dimensional (and with finite energy), effectively mapping the oscillator into an $N_{\rm max}$-dimensional qudit. In this setting, it has been established how to perform universal qudit-based quantum computation using oscillators~\cite{qudit2013mischuck,QuditsfromOscillatorsPhysRevA.104.032605}, yet an analogue of the Solovay-Kitaev theorem for oscillators, which can establish efficient approximation of arbitrary oscillator operations using \emph{discrete-parametrized} oscillator gates, is not known.

In general however, methods for performing CV computations by encoding continuous variables in DV devices (i.e., qudits) typically incur additional resource overhead compared to pure CV approaches.
For example, Ref.~\cite{C2QA-LGTpaper} introduces a method for simulating fermion-boson Hamiltonians on hybrid CV-DV quantum architectures, and shows that a purely DV-based approach incurs an increase in gate count of $O(\log^2(N_{\text{max}}))$ relative to the hybrid CV-DV method. Similar advantages of CV and hybrid approaches over purely DV methods are discussed in Refs.~\cite{Kumar_2025, sawaya2020resource}.

%

\section{Mixed Analog-Digital Quantum Signal Processing}
\label{sec:hybrid-qsp}

Leveraging the gates and operations defined in the previous section, here we present \emph{analog-digital quantum signal processing} for hybrid CV-DV systems. We first discuss hybrid single-variable QSP in Sec.~\ref{ssec:single_var_hybrid_qsp}, followed by a generalization to bi-variate non-Abelian QSP in Sec.~\ref{ssec:bivariable-qsp}. 

\subsection{Hybrid Single-variable QSP}\label{ssec:single_var_hybrid_qsp}
After analyzing the states and gates of qubits and oscillators, a natural question is how to compose these gates into useful quantum computations. This has been well-studied in DV quantum computation, leading to a variety of qubit-based algorithms~\cite{dalzell2023quantum}. 
Many of these DV algorithms admit adaptations to hybrid CV-DV quantum processors, as explicated in Refs.~\cite{andersen2015hybrid,liu2024hybrid}. As the focus of our work is signal processing, let us present CV-DV QSP~\cite{qspi2023}, as an adaptation of QSP to hybrid CV-DV processors.

As we review in Supplemental Material (SM)~Sec. SM.I, QSP provides a systematic framework for implementing a polynomial transformation of a linear operator, that is encoded in a block of a matrix (e.g., as one of its matrix elements). This is achieved by designing an alternating sequence of a fixed $z$-rotation that encodes the operator (i.e. the ``signal"), and parameterizable $x$-rotations. Such a \emph{QSP sequence} of length $d$ generates a degree $d$ polynomial, parameterized by the angles of the $x$-rotations. Importantly, for any polynomial bounded as $|P(x)|\leq 1$ over $-1\leq x \leq 1$, we can efficiently compute the corresponding angles with a classical algorithm~\cite{ chao2020finding,ying2022stable}. In fact, an early such method for determining these angles relied on Remez-type exchange algorithms~\cite{Fraser_1965}, ubiquitous in filter design in classical signal processing, thus inspiring the name ``quantum signal processing"~\cite{low2016methodology}.  

In generalizing QSP to hybrid CV-DV systems, let us define an important oscillator gate, the displacement gate $D(\alpha)$:
\begin{align}
    D(\alpha) 
    = e^{\alpha a^\dagger - \alpha^* a} = e^{i\sqrt{2}\Im{\alpha} \hat{x} - i\sqrt{2} \Re{\alpha} \hat{p}}
    \label{eq:displacement-gate}
\end{align}
where $\alpha \in \mathbb{C}$ is a complex number. As its name suggests, $D(\alpha)$ displaces the oscillator quadrature operators--- $\hat{x}$ by the real component of $\alpha$, and $\hat{p}$ by the imaginary component:
\begin{align}
    D^\dagger(\alpha) \, \hat{x} \, D(\alpha) = \hat{x} + \sqrt{2} \Re{\alpha} , \\
    D^\dagger(\alpha)\, \hat{p} \, D(\alpha) = \hat{p} + \sqrt{2} \Im{\alpha}.
    \label{eq:displacement-operator-action}
\end{align}
The displacement gate itself is considered a ``Gaussian" operation, meaning that it only manipulates the oscillator wave function classically and cannot generate non-classical states.
However, by coupling an oscillator to a qubit, the following entangling gate, known as the conditional displacement operation, can be generated:
\begin{align}
    W_z^{(\kappa)} = e^{-i\frac{\kappa}{2} \hat{x}  \sigma_z} = 
    \begin{bmatrix}
        \hat{w} & 0\\
        0 & \hat{w}^{-1}
    \end{bmatrix}, ~~~ ~~~ \hat{w} = e^{-i\frac{\kappa}{2}\hat{x}} . 
    \label{wz_k}
\end{align}
where $\kappa \in \mathbb{R}$ is a real-valued displacement parameter. This is a powerful operation that imparts a qubit-dependent momentum boost to the oscillator and can generate quantum entanglement between qubits and oscillators \cite{EickbuschECD}. The gate in Eq.\eqref{wz_k} is also a standard gate on trapped ion quantum computers \cite{HomeCharacteristicFunction}.
Importantly, we can equivalently interpret this as a $z$-rotation of the qubit through an angle $\kappa \hat x$ proportional to the position of the oscillator.
Likewise, Eq.~\eqref{wz_k} indicates that $W_z^{(\kappa)}$ encodes the variable $\hat{w}$ in its upper left block (of the $2 \times 2$ qubit subspace).
This identification suggests that $W_z^{(\kappa)}$ could be combined with parameterizable $\hat{x}$-rotations to develop a QSP sequence that generates a polynomial transformation of $\hat{w}$.

Following this intuition, we propose the following \emph{hybrid single-variable QSP} sequence of $d$ operations to produce a polynomial transformation in $\hat{w}$:
\begin{align}
    e^{i\phi_0 \sigma_x} \prod_{j = 1}^d W_z^{(\kappa)} e^{i\phi_j \sigma_x} =
    \begin{bmatrix}
        F(\hat{w}) & iG(\hat{w}) \\
        iG(\hat{w}^{-1}) & F(\hat{w}^{-1})
    \end{bmatrix}.
    \label{eq:qsp_wz}
\end{align}
This sequence is a natural extension of the ordinary QSP sequence (see Sec.~SM.I) to CV systems in order to build polynomials in the continuous variable $\hat{x}$. 
As we show in Sec.~SM.I, $F(\hat{w})$ and $G(\hat{w})$ are degree $d$ Laurent polynomials in $\hat{w}, \hat{w}^{-1}$ with real coefficients and parity $d \mod 2$ (i.e., consisting of only even or odd coefficients):
\begin{align}
    F(\hat{w}) &= \sum_{n = -d}^d f_n \hat{w}^n = \sum_{n = -d}^d f_n e^{-i\frac{n \kappa}{2}\hat{x}} := f(\hat{x}), \label{fwgw-poly_0} \\
    G(\hat{w}) &= \sum_{n = -d}^d g_n \hat{w}^n = \sum_{n = -d}^d g_n e^{-i\frac{n \kappa}{2}\hat{x}} := g(\hat{x})
    \label{fwgw-poly}
\end{align} 
where $f_n, g_n \in \mathbb{R}$. Evidently, $F(\hat{w})$ and $G(\hat{w})$ equate to periodic functions $f(\hat{x})$ and $g(\hat{x})$, both with periods of $\frac{4\pi}{\kappa}$ in position space. The coefficients $f_n, g_n$ can be computed by evaluating the Fourier series coefficients of $f(x)$ and $g(x)$ 
\begin{align}
    f_n = \sum_{n=-d}^d f(x) e^{i\frac{n \kappa}{2} x}, ~~g_n = \sum_{n=-d}^d g(x) e^{i\frac{n \kappa}{2} x} .
\end{align}
In addition, the unitarity of Eq.~\eqref{eq:qsp_wz} requires $F(\hat{w}) F(\hat{w}^{-1}) + G(\hat{w}) G(\hat{w}^{-1}) = I$.

Note that even though $\kappa\hat x$ is a quantum operator rather than a scalar rotation angle, this construction is identical to ordinary QSP as we discuss in Sec. SM.I, and thus the associated results carry over.
Crucially, for any real Laurent polynomial $F(\hat{w})$ as in Eq.~\eqref{fwgw-poly_0}, there exist phases $\{ \phi_0, \phi_1, ..., \phi_d \}$ such that the gate sequence of Eq.~\eqref{eq:qsp_wz} produces $F(w)$, and we can determine these phases. 

Moreover, the construction in Eq.~\eqref{eq:qsp_wz} can be changed to be a function of momentum $\hat p$
by instead using the following qubit-dependent position kick in place of $W_z^{(\kappa)}(\hat{\theta})$:
\begin{align}
    W_z^{(\lambda)} = e^{-i\frac{\lambda}{2} \hat{p} \sigma_z} &= 
    \begin{bmatrix}
        \hat{v} & 0\\
        0 & \hat{v}^{-1}
    \end{bmatrix}, ~~~ \hat{v} = e^{-i\frac{\lambda}{2}\hat{p}}.
    \label{wz_lambda}
\end{align}
More generally, to construct an operator that is a function of a linear combination of $\hat{x}$ and $\hat{p}$, such as $\frac{\kappa}{2} \hat{x} + \frac{\lambda}{2} \hat{p}$, one can apply QSP to the operator $e^{-i (\frac{\kappa}{2} \hat{x} + \frac{\lambda}{2} \hat{p} )\sigma_z}$. By varying the parameters $\kappa$ and $\lambda$, it is therefore possible to cover the entire phase space.
As such, this simple generalization of QSP to hybrid CV-DV systems allows us to implement a large class of operations on oscillators, with precision that improves with increasing polynomial degree $d$. This is useful in variety of applications; for example, Ref.~\cite{qspi2023} uses this construction to design an interferometer for quantum sensing applications in the few-shot limit, and Ref. \cite{liu2024hybrid} demonstrates how to use this technique to create a cat state in the oscillator by applying a conditional displacement followed by a QSP sequence.

\subsection{Hybrid Non-Abelian QSP}
\label{ssec:bivariable-qsp}
The hybrid single-variable QSP of the previous section is limited in application to functions of either $\hat{x}$, $\hat{p}$, or a linear combination thereof. Here we extend this constructing to multivariate functions in $\hat{x}$ and $\hat{p}$ by presenting \emph{hybrid non-Abelian QSP}. In a system comprised of one qubit and one oscillator, we define non-Abelian QSP by the following sequence of the two conditional displacements Eqs.~\eqref{wz_k} and~\eqref{wz_lambda}, interspersed with $X$ rotations parameterized by a set of phases $\{ \phi_{j}^{(k)}, \phi_{j}^{(\lambda)} \}$ for $j = 1,2,\cdots,d$:
\begin{align}
    U_d 
    &= e^{i\phi_0 \sigma_x} \prod_{j = 1}^d W_z^{(k)} e^{i\phi_j^{(k)} \sigma_x} W_z^{(\lambda)} e^{i\phi_j^{(\lambda)} \sigma_x} \nonumber \\
    &=
    \begin{bmatrix}
        F_d(w, v) & iG_d(w, v) \\
        iG_d(v^{-1}, w^{-1}) & F_d(v^{-1}, w^{-1})
    \end{bmatrix}.
    \label{qsp_u_bivar}
\end{align}
This sequence can be seen as a generalization of the recent multi-variable QSP sequence \cite{rossi2021mqsp} to position and momentum variables in the CV setting.
As we show in Sec. SM.II, this sequence implements a bivariate Laurent polynomial transformation in the non-commuting variables $\hat{w}$ and $\hat{v}$, which takes the following form:
\begin{align}
    F_d(\hat{w}, \hat{v}) = \sum_{r,s=-d}^d f_{rs} \hat{w}^r \hat{v}^s, ~~
    G_d(\hat{w}, \hat{v}) = \sum_{r,s=-d}^d g_{rs} \hat{w}^r \hat{v}^s,
    \label{f-g_coeff}
\end{align}
where $f_{rs}$ and $g_{rs}$ are complex coefficients parameterized by the phase angles $\{ \phi_j^{(k)}, \phi_j^{(\lambda)} \}$. Note that because $\hat{w}$ and $\hat{v}$ do not commute, their order in Eq.~\eqref{f-g_coeff} matters. Here we will express these polynomials with the factors of $\hat{w}$ always written to the left of $\hat{v}$, and refer to this convention as \emph{canonical}.

In using this construction, it would be desirable to show that for an arbitrary Laurent polynomial $F_d(\hat{w}, \hat{v})$, there always exist corresponding QSP phases $\{ \phi_{j}^{(k)}, \phi_{j}^{(\lambda)} \}$. However, the bi-variate nature of non-Abelian QSP renders the (single-variable) QSP theorem of Ref.~\cite{low2017optimal} inapplicable. Similarly, the non-commutativity of $\hat{w}$ and $\hat{v}$ inhibits application of the recent developments in multivariate QSP of commuting variables~\cite{rossi2021mqsp}. Accordingly, a characterization of the transformations achievable by non-Abelian QSP requires a more complete theory of polynomial transformations of two non-commuting variables, which has yet to be developed.


\section{Quantum AD/DA Conversion: Sampling and Interpolation of Quantum Data}
\label{sec:state-transfer-sampling}
\noindent

In this section, we present two methods for AD/DA conversion of \emph{quantum signals} on hybrid CV-DV systems. Analogous to classical signal processing, this procedure effectively realizes sampling and interpolation of quantum data. Physically, sampling transfers a CV state to a DV state, whereas interpolation transfers a DV state to a CV state. We will refer to these procedures as quantum A/D (analog-to-digital) and D/A (digital-to-analog) conversion, respectively. In the following, we will use subscripts $Q$ and $O$ to distinguish DV states (qubits) and CV states (oscillator), respectively.

Formally, in quantum D/A conversion, we begin with an $n$-qubit state $|\psi \rangle_Q = \sum_{\mathbf{x}} c_{\mathbf{x}} |\mathbf{x}\rangle_Q $, where $\mathbf{x} = (x_1, x_2, ..., x_{n})$ is a Boolean-valued vector and $|\mathbf{x}\rangle_Q = |x_1\rangle |x_2\rangle ... |x_{n} \rangle$ is the corresponding qubit state. We will also use the binary representation of integers, in which $\mathbf{x}$ corresponds to the integer $x= \sum_{j=1}^n x_j\cdot 2^{n-j} = x_1 \cdot 2^{n-1} + x_2 \cdot 2^{n-2} ... + x_n\cdot 2^0$. 
Our goal is to transfer $|\psi \rangle_Q$ to an analogous oscillator state $|\psi \rangle_O = \sum_{\mathbf{x}} c_{\mathbf{x}} |  x, \Delta \rangle_O $, where $| x ,  \Delta \rangle_O $ is a basis state in continuous space parameterized by the integer $x$ and a \emph{spacing parameter} $\Delta$. Intuitively, $| x ,  \Delta \rangle_O $ can be viewed as a state localized around the position $q=x \Delta$, such that adjacent basis states (i.e., $| x \pm 1 , \Delta \rangle_O $) are separated by $\Delta$; the exact form of the basis states depends on the AD/DA conversion implementation, as discussed below. 

Ultimately, we wish to construct an $n$-qubit \textit{quantum D/A conversion unitary} $U_{\text{D/A}}(\Delta, n)$, that obeys
\begin{align}\label{eq:ST_unitary}
    U_{\text{D/A}}(\Delta, n ) |\psi \rangle_Q |0, \Delta \rangle_O = |\mathbf{0}\rangle_Q | \psi \rangle_O,
\end{align}
where $|0, \Delta \rangle_O$ is the initial oscillator state, and $\mathbf{0} = [0,0,...,0]^T$. Crucially, the initial and final states must be unentangled to ensure that the information content of the initial state is fully transferred to the oscillator. Any residual entanglement would indicate that information remains in the quantum correlations between the systems, inaccessible to either party alone. Moreover, the inverse of this conversion unitary furnishes an $n$-qubit \textit{quantum A/D conversion unitary}:
\begin{align}
    U_{\text{D/A}}(\Delta, n )^\dag |\mathbf{0}\rangle_Q | \psi \rangle_O = |\psi \rangle_Q |0, \Delta \rangle_O . 
\end{align}

Below, we will present two methods to realize quantum AD/DA conversion. The first protocol uses hybrid single-variable QSP; the second protocol is an adaptation of the state transfer protocol of Ref.~\cite{hastrup2021improved}, which we show is an instance of non-Abelian QSP. For both protocols, we prove analytical bounds on their performance and resource requirements.

\subsection{Quantum AD/DA Conversion: Hybrid Single-Variable QSP}\label{ssec:ST_single_variable_QSP}
\begin{figure}[tbp]
    \begin{center}
    \includegraphics[width=0.995\linewidth]{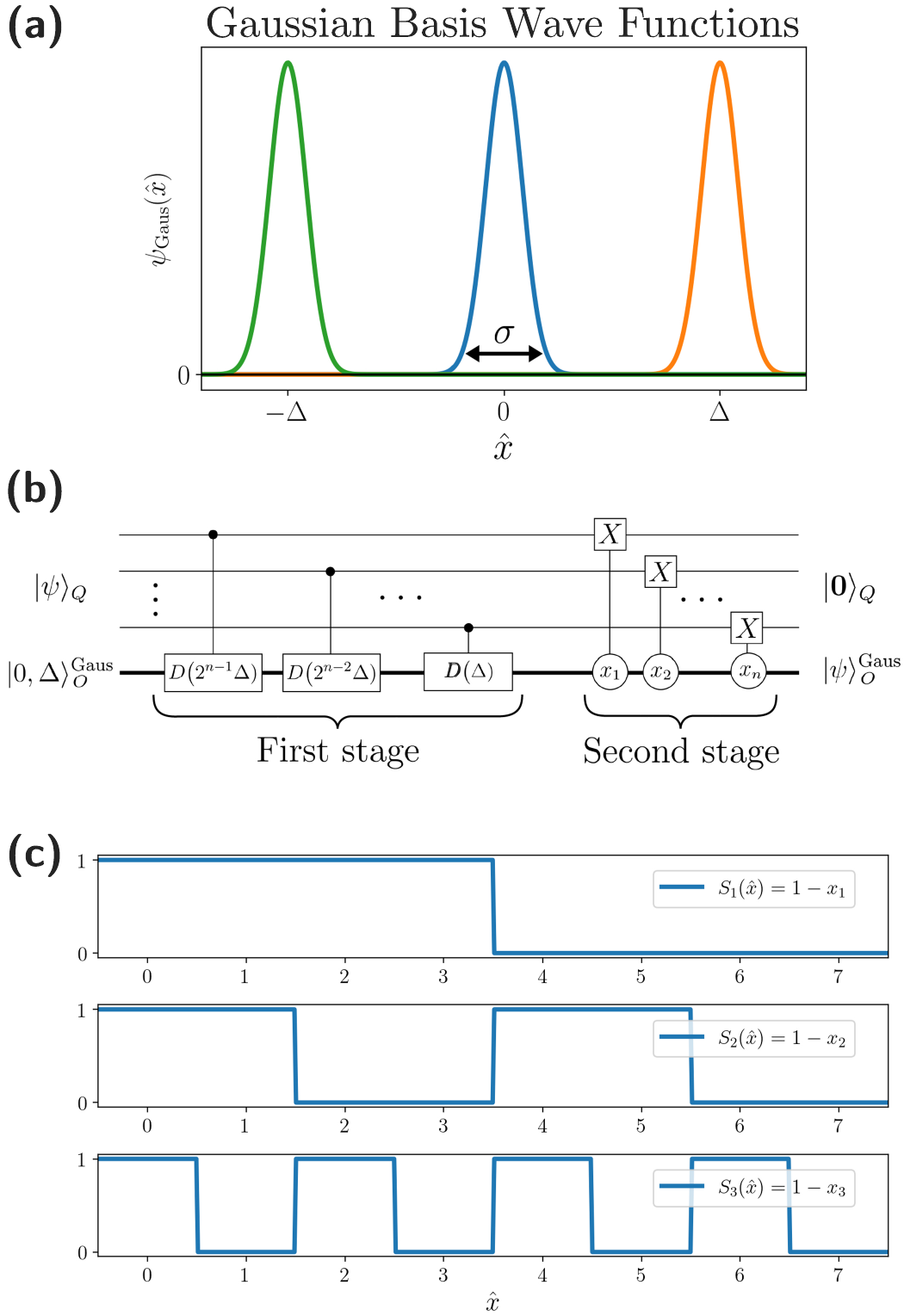}
    \end{center}
    \caption{ \textbf{(a):} Illustration of the Gaussian basis states used in D/A conversion with single-variable QSP. The wave functions in position space are Gaussians of width $\sigma$, each separated by the spacing parameter $\Delta$. \\
    \textbf{(b):} The quantum circuit that implements D/A conversion with single variable QSP.
    Here, thin lines denote qubits, and the thick line an oscillator. Time proceeds left to right, enacting the gates depicted as boxes.
    The initial qubits state is $|\psi\rangle_Q$, and the initial oscillator state is $|0,\Delta\rangle_O^{\text{Gaus}}$. The first stage applies a series of controlled displacements $D( 2^{n-j} \Delta)$ between the qubits and oscillator. The second stage applies a series of operations that disentangle the qubits by flipping qubit $j$ conditioned on the bit $x_j$ of the oscillator's position. We depict these operations as an $X:= \sigma_x$ gate conditioned on $x_j$. In practice, each of these is realized as a QSP sequence $R_j(\hat{x})$ according to the construction of Eq.~\eqref{eq:W_j_QSP}. \\
    \textbf{(c):} The square wave functions $S_j(x\Delta)$ of Eq.~\eqref{eq:square_wave_func} for $j=1,2,3$, with $n=3$ and $\Delta=1$. Observe how at integer values $x$, these square waves are equal to $1-x_j$, which enables one to read out the bits $\{ x_j \}$.
    }\label{fig:QSP_state_transfer}
\end{figure}

Our first quantum AD/DA conversion protocol employs hybrid single-variable QSP. For this protocol, it would be ideal for the oscillator basis states $|x , \Delta \rangle_O$ to be infinitely localized, which would simplify analysis and the transfer of states. Equivalently, this would correspond to a position eigenstate $|q\rangle_O$, with eigenvalues $q=x\Delta$ at integer multiples of the spacing parameter. Realistically however, an exact position eigenstate is un-normaliazable and cannot be prepared, so we instead take our basis states to be Gaussians of width $\sigma \ll \Delta$ centered around $x \Delta$: 
\begin{equation}\label{eq:ST_Gaussian_basis_QSP}
    |x, \Delta \rangle_O^{\text{Gaus}} := \frac{1}{\sqrt{\sigma} (2\pi)^{1/4}} \int dq e^{-(q-x\Delta)^2/4\sigma^2} |q\rangle_O,
\end{equation}
which reduces to a position eigenstate as $\sigma \rightarrow 0$. These states are nearly orthonormal for small $\sigma/\Delta$:
\begin{equation}\label{eq:stateTransIP}
    ^{\text{Gaus}}_{ \ \ O}\langle y, \Delta  |x, \Delta \rangle_O^{\text{Gaus}} = e^{-\frac{\Delta^2}{\sigma^2} \frac{(x-y)^2}{8}},
\end{equation}
which approaches the Kronecker delta $\delta_{x,y}$ as $\sigma/\Delta \rightarrow 0$. For visual intuition, we plot examples of thees wave functions in Fig.~\ref{fig:QSP_state_transfer}a.

Gaussian states are standard in CV quantum computing, as they correspond to the ground state of the quantum harmonic oscillator, and can be prepared by either cooling a system to its ground state or projecting onto the ground state with through measurements. This is standard procedure across various platforms, including microwave resonators, trapped ions, and neutral atoms. Likewise, the standard deviation of a Gaussian state can be tuned through single-mode squeezing, and the mean can be adjusted by applying a displacement gate~\cite{liu2024hybrid}.

\textbf{D/A Conversion Protocol:} We first focus on the D/A conversion protocol, as the corresponding A/D protocol is simply its inverse. The D/A protocol consists of two stages: first, a series of conditional displacements are applied between the qubits and oscillator, and then the intermediate state is disentangled by a series of hybrid single-variable QSP operations. We illustrate the circuit of the D/A conversion protocol in Fig.~\ref{fig:QSP_state_transfer}b, showcasing its decomposition into conditional displacements and QSP operations. We will denote the corresponding D/A conversion unitary by $U_{\text{D/A}}^{\text{S-V}}(\Delta,n)$, using ``S-V" for single-variable QSP.

\textit{Stage 1 -- Displacement:} First, we apply a series of controlled displacements to the oscillator: $\prod_{j=1}^{n} D_{j}(\Delta 2^{n-j})$,
where $D_{j}(\Delta 2^{n-j})$ is a displacement (as in Eq.~\eqref{eq:displacement-gate}), controlled by the $j^{\text{th}}$ qubit. This transforms the initial state as
\begin{equation}\label{eq:QSP1_intermediate_state}
\begin{aligned}
   |\psi\rangle_Q |0, \Delta \rangle_O^{\text{Gaus}} &= \sum_{\mathbf{x}} c_{\mathbf{x}} |\mathbf{x}\rangle_Q |0, \Delta \rangle_O^{\text{Gaus}} \\
   & \qquad \qquad \mapsto \sum_{\mathbf{x}} c_{\mathbf{x}} |\mathbf{x}\rangle_Q |x , \Delta \rangle_O^{\text{Gaus}}.
\end{aligned}
\end{equation}
We could imagine disentangling this state as 
\begin{equation}\label{eq:intermediate_state_map}
\sum_{\mathbf{x}} c_{\mathbf{x}} |\mathbf{x}\rangle_Q |x , \Delta \rangle_O^{\text{Gaus}} \mapsto \sum_{\mathbf{x}} c_{\mathbf{x}} |\mathbf{0}\rangle_Q |x , \Delta \rangle_O^{\text{Gaus}} = |\mathbf{0}\rangle_Q | \psi \rangle_O^{\text{Gaus}}, 
\end{equation}
to achieve the desired D/A conversion, where $| \psi \rangle_O^{\text{Gaus}}$ is the transferred wave function encoded in the basis of Gaussian states. We realize this disentangling procedure with hybrid single-variable QSP as follows.

\textit{Stage 2 -- QSP:} In the QSP stage, we disentangle the qubits and oscillator by enacting an operation that sets each qubit to $|0\rangle$ as in Eq.~\eqref{eq:intermediate_state_map}. This operation would act as $|x_j \rangle |x, \Delta \rangle_O^{\text{Gaus}} \mapsto |0 \rangle |x,\Delta \rangle_O^{\text{Gaus}}$ for each qubit $j$, or equivalently flip qubit $j$ conditioned on the bit $x_j$. Here we implement this operation as a series of hybrid single-variable QSP sequences, one for each qubit.

This desired behavior requires that we determine the bits $\{ x_j \}$ from the oscillator's binary representation $| x , \Delta \rangle_O^{\text{Gaus}}$. Observe that the bit $x_j$ of the position $x \Delta = \Delta \sum_{j=1}^n 2^{n-j} x_j $ can be read out by a ``square wave" function: 
\begin{equation}\label{eq:square_wave_func}
    S_j(\hat{x}) := \Theta \left(\cos\left[\tfrac{\pi}{2^{n-j}} \left(\tfrac{\hat{x}}{\Delta} - 2^{n-j-1}+\tfrac{1}{2}\right)\right] \right) = 1-x_j,
\end{equation}
where $\Theta(\cdot)$ is the Heaviside step function. For visual intuition, we depict this function in Fig.~\ref{fig:QSP_state_transfer}c, illustrating how such a square wave function outputs the bits $x_j$.

We may then use this observation to construct the unitary
\begin{align}\label{eq:W_j_QSP}
    R_j(\hat{x}) :&= 
    \begin{pmatrix}
        S_j(\hat{x}) & \sqrt{1-S_j(\hat{x})^2} \\
        -\sqrt{1-S_j(\hat{x})^2} & S_j(\hat{x})
    \end{pmatrix}\nonumber\\
    &= 
    \begin{cases}
    I & x_j = 0 \\
    i\sigma_y & x_j = 1,
    \end{cases}
\end{align}
where $\hat{x}$ denotes the position operator. This operation correctly flips the $j^{\text{th}}$ qubit conditioned on $x_j$: $R_j(\hat{x}) |x_j\rangle | x , \Delta \rangle_O^{\text{Gaus}} = |0\rangle | x,  \Delta \rangle_O^{\text{Gaus}}$.\footnote{Here we take $|x , \Delta \rangle_O^{\text{Gaus}}$ to be an exact position eigenstate; we will remedy this assumption and evaluate performance on $|x,\Delta \rangle_O^{\text{Gaus}}$ shortly.} Therefore, the sequence $\prod_{j=1}^{n} R_j(\hat{x})$ correctly disentangles all $n$ qubits.

Our strategy is to approximate each $R_j(\hat{x})$ as a hybrid single-variable QSP sequence. In particular, for the $j$th sequence, we will choose our variable to be $\hat{w}_j = e^{\frac{\pi}{2^{n-j}}(\hat{x}/\Delta - 2^{n-j-1}+1/2)}$. We can then employ hybrid single-variable QSP to implement a real-valued Laurent polynomial that approximates the step function as $F(\hat{w}_j) \approx S_j(\hat{x})$, corresponding to the operation (see Eq.~\eqref{eq:qsp_wz}):
\begin{equation}
    \begin{pmatrix}
        F(\hat{w}_j) & i\sqrt{1-F(\hat{w}_j)^2} \\
        i\sqrt{1-F(\hat{w}_j)^2} & F(\hat{w}_j)
    \end{pmatrix}.
\end{equation}
This follows from choosing $F(\hat{w}_j)$ to be real, as $F(\hat{w}_j) = F(\hat{w}_j^{-1}) \in \mathbb{R}$. 
Upon conjugation by a phase gate $S$, this operation becomes
\begin{equation}
\begin{aligned}
    & \begin{pmatrix}
        F(\hat{w}_j) & \sqrt{1-F(\hat{w}_j)^2} \\
        -\sqrt{1-F(\hat{w}_j)^2} & F(\hat{w}_j)
    \end{pmatrix} =: \tilde{R}_j(\hat{x}),
\end{aligned}
\end{equation}
which approximates $\tilde{R}_j(\hat{x}) \approx R_j(\hat{x})$ because $F(\hat{w}_j) \approx S_j(\hat{x})$, where the accuracy in this approximation is dictated by the polynomial approximation. The accuracy and cost of such an approximation is established in the literature: a QSP polynomial can approximate the step function to within some error $\epsilon$, except within a region of width $\delta$ centered around the discontinuity, and the degree of this polynomial is $\mathcal{O} \left( \frac{1}{\delta} \log(\frac{1}{\epsilon}) \right)$~\cite{low2017hamiltonian,martyn2023efficient}. In aggregate then, by applying the series of QSP sequences $\prod_{j=1}^n \tilde{R}_j(\hat{x})$ to intermediate state of Eq.~\eqref{eq:QSP1_intermediate_state}, we disentangle the qubits and oscillator, and (approximately) produce the desired final state $|\mathbf{0}\rangle_Q |\psi\rangle_O $. 

\textbf{A/D Conversion Protocol}: Because the D/A conversion protocol is unitary, its inverse furnishes an analogous quantum A/D conversion protocol. In this direction, take the initial state to be $| \mathbf{0}\rangle_Q |\psi\rangle_O^{\text{Gaus}} =| \mathbf{0}\rangle_Q \sum_x c_x |x,\Delta\rangle_O^{\text{Gaus}}$. Then, by applying the inverted sequence $\prod_{j=n}^1 \tilde{W}_j(\hat{x})^\dag$, and subsequently the inverted controlled displacements $\prod_{j=n}^1 D_j(-\Delta 2^{n-j})$, one obtains the transferred state $|\psi\rangle_Q | 0, \Delta \rangle_O^{\text{Gaus}} $. 

\begin{figure}[tbp]
    \centering
    \includegraphics[width=\columnwidth]{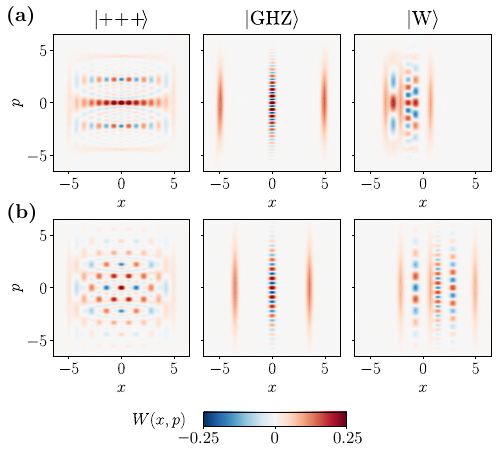}
    \caption{D/A conversion for various three-qubit states using $\textbf{(a)}$ single-variable QSP and $\textbf{(b)}$ non-Abelian QSP, including $\ket{\textrm{GHZ}}=(\ket{000}+\ket{111})/\sqrt{2}$ and $\ket{\textrm{W}}=(\ket{001}+\ket{010}+\ket{100})/\sqrt{3}$. 
    $\textbf{(a)}$: We use a single-variable QSP sequence of degree $d=60$ with $\delta=0.2$, $\Delta=1$, and $\sigma = e^{-1.12} \approx 0.37$. As a metric for successful conversion, we estimate the purity (a measure of the degree to which the oscillator and qubits have been successfully disentangled) of the final oscillator state, yielding $0.976$, $0.958$, and $0.982$, respectively. These simulations were carried out in Bosonic Qiskit~\cite{Biskit}.
    $\textbf{(b)}$: We use non-Abelian QSP with $\Delta = \sqrt{2}$ and approximate the initial oscillator sinc state (defined in Eq.~\eqref{eq:sinc_state}) by a Gaussian with $\sigma = e^{-1.12} \approx 0.37$. The purities of the final oscillator state evaluate to $0.858$, $0.858$, and $0.858$, respectively. We used QuTiP \cite{johansson2012qutip} for these simulations.}
    \label{fig_statetransfer_examples_new}
\end{figure}

\textbf{Performance:} The circuit of the D/A conversion protocol is illustrated in Fig.~\ref{fig:QSP_state_transfer}b. To provide visual intuition on this protocol, we also depict in Fig.~\ref{fig_statetransfer_examples_new}a the Wigner function (e.g. a phase space quasiprobability distribution; see Ref. \cite{liu2024hybrid} for a detailed definition) of the oscillator upon D/A conversion for various initial 3-qubit states. Let us now analyze the gate complexity and error of this protocol.

The first stage requires $n$ displacement gates of sizes $\Delta 2^{n-j}$ for $j=1,..., n$. This translates to a total displacement amount 
\begin{equation}
    \sum_{j=1}^{n} \Delta 2^{n-j}= \mathcal{O}(\Delta 2^n),
\end{equation}
and thus a time complexity $\mathcal{O}(\Delta 2^n)$. This scales as $2^n$ when the controlled displacements are implemented with a fixed coupling between the qubits and oscillator, yet can be reduced if sufficient squeezing is available on the quantum device (e.g. by selecting $\Delta = \mathcal{O}(2^{-n})$).

In the second stage, we take the polynomial implemented by the $j^{\text{th}}$ QSP sequence to be an approximation to the step function that suffers error at most $\epsilon$ outside of a region of width $\delta_j$ centered about the discontinuity. Each such QSP sequence requires $\mathcal{O}(\frac{1}{\delta_j}\log(1/\epsilon))$ gates~\cite{low2017optimal,GrandUnificationAlgos}. To ensure that the $j^{\text{th}}$ QSP sequence can discern the correct bit $x_j$ when acting on a state located at position $q=x\Delta$, we require that the width of the approximate step function be $\delta_j \leq \mathcal{O}(\tfrac{1}{2^{n-j}})$. Therefore, the total gate complexity of all the QSP sequences is 
\begin{align}
    \mathcal{O}\big( \textstyle \sum_j \tfrac{1}{\delta_j}\log(1/\epsilon)\big) =\mathcal{O}(2^n \log(1/\epsilon)).
\end{align} 
As the gates comprising the QSP sequences (i.e. rotations) take time $\mathcal{O}(1)$, this corresponds to a time complexity $\mathcal{O}(2^n \log(1/\epsilon))$. 
The total time complexity of this D/A conversion is thus
\begin{align}
    T_{\rm D/A} = \mathcal{O}\left(2^n\left(\Delta + \log(1/\epsilon)\right)\right).
\end{align}

Lastly, to analyze the fidelity of this D/A conversion, note its two sources of error: first, the basis states are not exact position eigenstates but rather Gaussians of finite width $\sigma$; and second, the QSP polynomial is only an accurate approximation to within error $\epsilon$, and fails near the discontinuity of the step function. A careful analysis of these errors, presented in Sec. SM.III, indicates that the fidelity between the output state of this protocol and the desired state $\ket{\mathbf{0}}_Q \ket{\psi}_O^{\text{Gaus}}$ is 
\begin{equation}
    1-\mathcal{O}(n\epsilon) - e^{-\mathcal{O}(\Delta^2/\sigma^2)}.
\end{equation}
Note that this depends on the ratio $\sigma/\Delta$, and consequently, the degree to which the basis states are orthogonal.

Lastly, recall that the inverse of this protocol provides a quantum A/D conversion protocol. Because the inverse is just the time-reversed operation, its gate and time complexities are the same as that of D/A conversion, as is the asymptotic expression for the fidelity.

Collectively, the results of these AD/DA conversion protocols can be summarized as follows:
\begin{restatable}[Quantum AD/DA Conversion with Hybrid Single Variable QSP]{theorem}{si} \label{thm:ST_single_var_QSP}
    The quantum D/A conversion protocol based on hybrid single-variable QSP achieves a fidelity $1-\mathcal{O}(n\epsilon) - e^{-\mathcal{O}(\Delta^2/\sigma^2)} \big) $, at a gate complexity of $\mathcal{O}(2^n\log(1/\epsilon))$ and time complexity $\mathcal{O}(2^n (\Delta + \log(1/\epsilon)))$, where $n$ is the number of qubits of the DV state, $\epsilon$ is the error on the the polynomial realized by QSP, $\sigma$ is the width of the initial Gaussian wave function of the oscillator, and $\Delta$ is a spacing parameter.

    Analogously, in reverse this furnishes a quantum A/D conversion protocol that achieves a fidelity $1-\mathcal{O}(n\epsilon) - e^{-\mathcal{O}(\Delta^2/\sigma^2)} $ and identical gate and time complexities. 
\end{restatable}

For constant $\Delta$, the runtime scales linearly in the dimension of the DV Hilbert space, as we alluded to in the introduction. This is expected as we are directly encoding the initial state in $2^n$ basis states equally spaced apart in position space. In principle, one could more efficiently encode this information via a binary encoding on multiple oscillators. 

Furthermore, the fidelity is maximized in the limit of small QSP error $\epsilon$, and a small ratio $\sigma/ \Delta$ (i.e., a large relative spacing between basis states). Therefore, one can improve fidelity by either squeezing the initial state to decrease $\sigma$, or selecting a larger spacing $\Delta$. To achieve a fidelity $1 - \varepsilon$, it suffices to select $\epsilon = \mathcal{O}(\varepsilon/n)$ and $(\Delta/\sigma)^2 = \mathcal{O}(\log(1/\varepsilon))$, translating to an overall time complexity
\begin{equation}
    T_{\rm D/A} = \mathcal{O}\left(2^n \left( \sigma \sqrt{\log(1/\varepsilon)} + \log(n/\varepsilon) \right) \right). 
\end{equation}

\subsection{Quantum AD/DA Conversion: Hybrid Non-Abelian QSP}\label{ssec:ST_Non_Abelian_QSP}
Recently, Ref.~\cite{hastrup2022universal} proposed a method to transfer a CV state to an $n$-qubit state by enacting a series of controlled displacements between the oscillator and each qubit. This naturally defines an A/D conversion protocol, which in reverse furnishes a D/A conversion protocol. Below, we review these protocols and provide bounds on their performance. We also show how these protocols can be viewed as instances of hybrid non-Abelian QSP, and thus we will refer to them accordingly.

\textbf{A/D Conversion Protocol:}
To begin, let us denote the A/D conversion unitary of Ref.~\cite{hastrup2022universal} by $U_{\text{D/A}}^{\text{N-A}}(\Delta, n)^\dag$, using ``N-A" for non-Abelian. We define this by its Hermitian conjugate such that its inverse $U_{\text{D/A}}^{\text{N-A}}(\Delta, n)$ performs D/A conversion, in line with our notation of Eq.~\eqref{eq:ST_unitary}. This operation uses a spacing parameter $\Delta$ and transfers a CV state to $n$ qubits.

The initial state of this protocol is $|\mathbf{0}\rangle_Q |\psi\rangle_O $, where $|\psi\rangle_O = \int dq \ \psi(q) dq |q\rangle_O$ is a state on the oscillator to be transferred to the $n$ qubits. Explicitly, $U_{\text{D/A}}^{\text{N-A}}(\Delta, n)^\dag$ is the unitary operation
\begin{equation}\label{eq:Hastrup_state_tr_operator}
    U_{\text{D/A}}^{\text{N-A}}(\Delta, n)^\dag = \prod_{j = n}^{1} W_j V_j = W_n V_n \cdots W_1 V_1 ,
\end{equation}
where 
\begin{equation}
    V_j = e^{i \frac{\pi}{2^{j} \Delta} \hat{x}\hat{\sigma}_y^{(j)}}, \quad 
    W_j = 
    \begin{cases} 
        e^{i \frac{\Delta}{2} 2^{j-1}\hat{p}\hat{\sigma}_x^{(j)}} & j<n, \\
        e^{-i \frac{\Delta}{2} 2^{j-1}\hat{p}\hat{\sigma}_x^{(j)}} & j=n,
    \end{cases}
\end{equation}
are momentum boosts and displacements of the oscillator, and the superscript ${(j)}$ denotes action on the $j^{\text{th}}$ qubit (e.g. $\sigma_x^{(j)}$ acts on qubit $j$). 
This sequence is carefully chosen such that the momentum boosts and displacements conspire together to map the CV wave function $\psi(q)$, evaluated at a discrete set of positions $q_{\mathbf{s}}$, onto the amplitudes of a DV quantum state, thus performing quantum A/D conversion.

In more detail, Ref.~\cite{hastrup2022universal} analyzes this sequence and shows that application of $U_{\text{D/A}}^{\text{N-A}}(\Delta, n)^\dag$ to the initial state $|\mathbf{0}\rangle_Q |\psi\rangle_O $ outputs the state
\begin{align}\label{eq:HastrupTransferedState}
&\sum_{\mathbf{s} \in \{-1, +1 \}^n} \int dq \ \psi(q+q_{\mathbf{s}}) \prod_{j=1}^n \cos(\tfrac{\pi q}{\Delta 2^j}) |\phi_{\mathbf{s}} \rangle_Q |q\rangle_O ,
\end{align}
where the sum runs over all $\mathbf{s} \in \{-1, +1 \}^n$, and the basis states are 
\begin{equation}\label{eq:phi_s}
    |\phi_{\textbf{s}} \rangle = (-1)^{\gamma_{\textbf{s}}} \cdot \bigotimes_{j=1}^n \left[ \frac{1}{\sqrt{2}} \big(|0\rangle +s_j |1\rangle \big) \right] ,
\end{equation} 
for a scalar $\gamma_s$ defined as
\begin{equation}
    \gamma_{\mathbf{s}} = \sum_{j=1}^{n-2} \frac{1}{2}(s_j + s_{j+1}) + \frac{1}{2}(s_{n-1} - s_n) .
\end{equation} 
For instance, if $\mathbf{s} = (1,-1,1,-1)$, then $\gamma_{\mathbf{s}} = 1$, and $|\phi_{\mathbf{s}} \rangle = - |+\rangle |-\rangle |+\rangle |-\rangle$. In addition, the value $q_{\textbf{s}}$ in Eq.~\eqref{eq:HastrupTransferedState} is 
\begin{equation}\label{eq:q_s}
    q_{\textbf{s}} = \frac{\Delta}{2} \Bigg( \sum_{j=1}^{n-1} s_j 2^{j-1} - s_n 2^{n-1} \Bigg).
\end{equation} 
This quantity takes $2^n$ discrete values in the range $[-\tfrac{\Delta}{2} (2^n-1), \tfrac{\Delta}{2} (2^n-1)]$, with each possible value equally spaced by $\Delta$.

Two approximations are used in Ref.~\cite{hastrup2022universal} to simplify the state of Eq.~(\ref{eq:HastrupTransferedState}). First, it is assumed that the support of $\psi(q)$ is limited to $|q| \leq \tfrac{\Delta}{2} (2^n-1)$, such that one can make the replacement $\prod_{j=1}^n \cos(\tfrac{\pi q}{\Delta 2^j}) \approx \text{sinc}(\tfrac{\pi q}{\Delta})$ over the support of the wave function. Second, it is also assumed that $\int dq \ \psi(q+q_{\mathbf{s}}) \text{sinc}(\tfrac{\pi q}{\Delta}) \approx \psi(q_{\mathbf{s}}) \int dq \  \text{sinc}(\tfrac{\pi q}{\Delta}) $, which dictates that $\psi(q)$ be slowly varying relative to $\text{sinc}(\tfrac{\pi q}{\Delta})$, i.e., $|\tfrac{d\psi}{dq}| \ll 1/\Delta$. With both of these approximations made, Eq.~\eqref{eq:HastrupTransferedState} simplifies to 
\begin{equation}\label{eq:HastrupTransferedState_Approx}
    \sum_{\mathbf{s}} \sqrt{\Delta} \psi(q_{\mathbf{s}}) |\phi_{\mathbf{s}} \rangle_Q  \otimes \frac{1}{\sqrt{\Delta}} \int dq \ \text{sinc}(\tfrac{\pi q}{\Delta}) |q\rangle_O . 
\end{equation}
Notably, the oscillator is now decoupled from the qubits, and therefore the initial CV state $ \psi(q) $ has been transferred to a corresponding qubits state $\sum_{\mathbf{s}} \sqrt{\Delta} \psi(q_{\mathbf{s}}) |\phi_{\mathbf{s}} \rangle_Q$, encoded in the $\{ |\phi_{\mathbf{s}} \rangle_Q \} $ basis.

\begin{figure}[tbp]
    \begin{center}
    \includegraphics[width=0.96\linewidth]{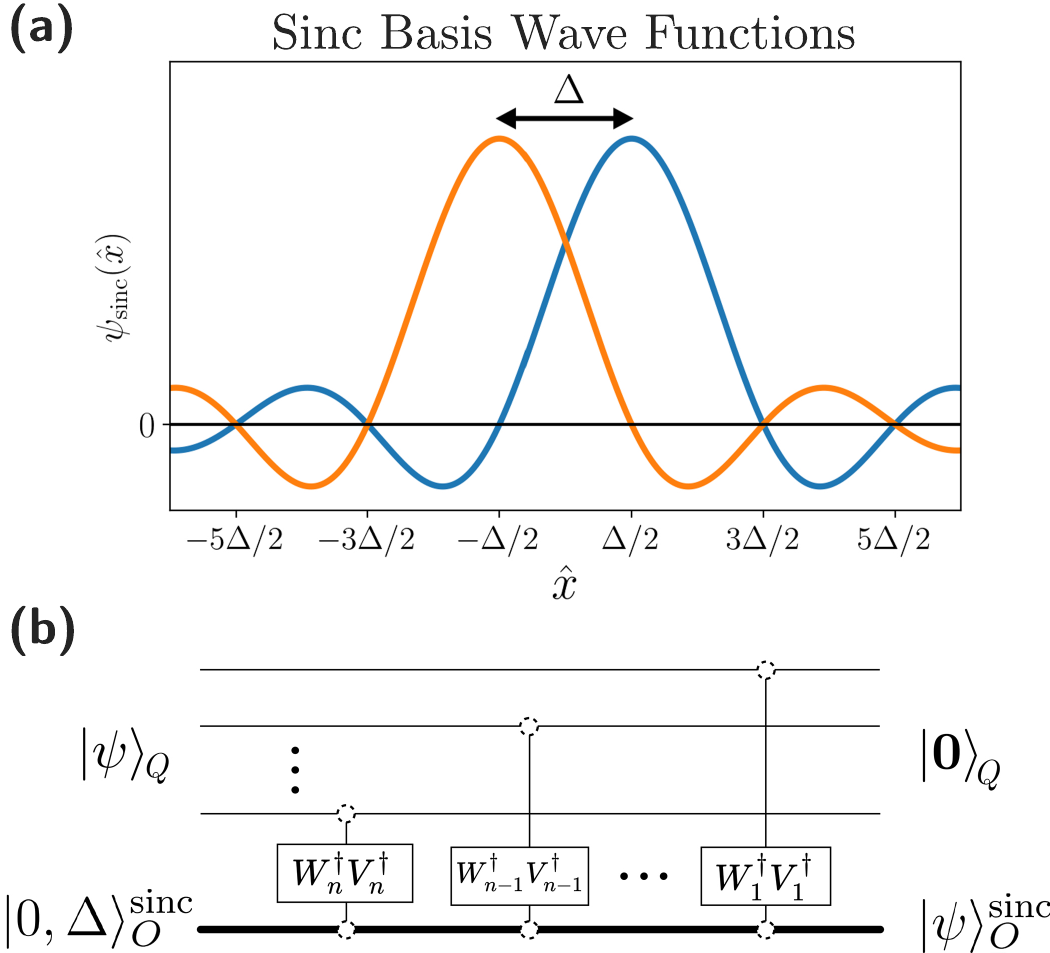}
    \end{center}
    \caption{\textbf{(a):} Illustration of two sinc basis states, as used in D/A conversion with non-Abelian QSP. The wave functions in position space are sinc functions (Eq.~\eqref{eq:sinc_state}), and have peaks that are each separated by the spacing parameter $\Delta$. \\
    \textbf{(b):} The circuit that implements D/A conversion with non-Abelian QSP, adapted from Ref.~\cite{hastrup2022universal} with the order of $W_n$ and $V_n$ flipped, which we believe to be a typo in Fig. 1 of Ref.~\cite{hastrup2022universal}. The initial qubits state is $|\psi\rangle_Q$, and the initial oscillator state is a sinc state $|0, \Delta \rangle_O^{\text{sinc}} = \frac{1}{\sqrt{\Delta}} \int dq \  \text{sinc}(\pi q /\Delta) |q\rangle_O$. Then, one applies a series of operations $V_j^\dag W_j^\dag $ between the oscillator and the $j^{\text{th}}$ qubit, where $V_j = e^{i \frac{\pi}{2^{j} \Delta} \hat{x}\hat{\sigma}_y^{(j)}}$ and $W_j = e^{ \pm i \frac{\Delta}{2} 2^{j-1}\hat{p}\hat{\sigma}_x^{(j)}}$. The systems on which these operations act are denoted by circles with dashed lines. In aggregate, this maps the initial qubits state to an equivalent oscillator state $|\psi\rangle_O$ encoded in a basis of displaced sinc states, as per Eq.~\eqref{eq:Hastrup_inverse_state_transfer}.}
    \label{fig:Hastrup_state_transfer}
\end{figure}

\textbf{D/A Conversion Protocol:}
This A/D conversion protocol can be run in reverse to achieve D/A conversion. In this direction, one first prepares the qubits in the state $\sum_{\mathbf{s}} c_{\mathbf{s}} |\phi_{\mathbf{s}} \rangle$, and the oscillator in the \emph{sinc state} $|0, \Delta\rangle_O^{\text{sinc}} = \frac{1}{\sqrt{\Delta}} \int dq \ \text{sinc}(\tfrac{\pi q}{\Delta}) |q\rangle_O$.\footnote{As explained in Ref.~\cite{hastrup2022universal}, this exact state is unphysical because it has infinite energy, but it can be well approximated by a squeezed vacuum.} Then, enacting $U_{\text{D/A}}^{\text{N-A}}(\Delta, n) = \prod_{j = 1}^{n} V_j^\dag W_j^\dag $ (approximately) outputs the state 
\begin{equation}\label{eq:Hastrup_inverse_state_transfer}
\begin{aligned}
    &|\mathbf{0}\rangle_Q \otimes \sum_{\mathbf{s}}  c_{\mathbf{s}} \frac{1}{\sqrt{\Delta}} \int dq \ \text{sinc} \left( \tfrac{\pi (q-q_{\mathbf{s}})}{\Delta} \right) |q\rangle_O = | \mathbf{0}\rangle_Q
    |\psi\rangle_O^{\text{sinc}}  .
\end{aligned}
\end{equation}
This has transferred the initial DV state to a CV state encoded in the basis of displaced ``sinc states":
\begin{equation}\label{eq:sinc_state}
    |q_{\mathbf{s}}, \Delta \rangle_O^{\text{sinc}} := \frac{1}{\sqrt{\Delta}} \int dq \ \text{sinc} \left( \tfrac{\pi (q-q_{\mathbf{s}})}{\Delta} \right) |q\rangle_O .
\end{equation}
A sinc state is a state in continuous space that is localized around $q=q_s$, with adjacent sinc state separated by $\Delta$, such they are orthonormal: $^{\text{sinc}}_O\langle q_{\mathbf{s}}, \Delta  | q_{\mathbf{s}'}, \Delta \rangle_O^{\text{sinc}} = \delta_{\mathbf{s} \mathbf{s}'}$. We illustrate examples of these basis states in Fig.~\ref{fig:Hastrup_state_transfer}a. Satisfyingly, this representation of $|\psi\rangle_O^{\text{sinc}}$ as a sum of displaced sinc functions is analogous to the construction of Shannon's sampling theorem. By this connection, this protocol can be viewed as a quantum realization of Shannon's sampling theorem.

\textbf{Recontextualization as Non-Abelian QSP:}
This AD/DA conversion protocol can be reinterpreted as an instance of non-Abelian QSP. In the D/A direction, we rewrite the term $V_j^\dag W_j^\dag$ in the language of non-Abelian QSP as
\begin{equation}\label{qft_readin-qsp}
\begin{aligned}
     & e^{-i\frac{\pi}{4} \sigma_y^{(j)}} (V_j^\dag W_j^\dag
       ) e^{i\frac{\pi}{4} \sigma_y^{(j)}}
    = e^{-i \frac{\pi}{\Delta 2^{j} } \hat{x}\hat{\sigma}_y^{(j)}} e^{\mp i \frac{\Delta}{2} 2^{j-1}\hat{p}\hat{\sigma}_z^{(j)}} \\
    =& e^{i\frac{\pi}{4} \sigma_x^{(j)}} e^{-i \frac{\pi}{\Delta 2^{j}} \hat{x}\hat{\sigma}_z^{(j)}}  e^{-i\frac{\pi}{4} \sigma_x^{(j)}} e^{\mp i \frac{\Delta}{2} 2^{j-1}\hat{p}\hat{\sigma}_z^{(j)}},
\end{aligned}
\end{equation}
where the $-$ sign is taken for $j<n$, and the $+$ sign for $j = n$. 
By comparing this expression to the non-Abelian QSP sequence of Eq.~\eqref{qsp_u_bivar}, it is readily identified that $V_j^\dag W_j^\dag$, upon conjugation by $e^{-i\frac{\pi}{4} \sigma_y^{(j)}}$, corresponds to a degree-1 non-Abelian QSP sequence with the displacement amounts
\begin{align}
    k = \frac{2 \pi}{\Delta 2^{j}},~~~~\lambda = \mp \frac{\Delta 2^j}{2} ,
\end{align}
for the $j^{\text{th}}$ qubit, and with QSP phases
\begin{align}
    \phi_0 = \frac{\pi}{4}, ~~~\phi_1^{(k)} = - \frac{\pi}{4}, ~~~\phi_1^{(\lambda)} = 0
\end{align}
for all $j$ qubits. In this incarnation, this AD/DA conversion protocol may be interpreted as a product of $n$ degree-$1$ non-Abelian QSP sequences, where each sequence acts between the oscillator and the $j^{\text{th}}$ qubit. This identification suggests that this protocol could admit a generalization by using a higher degree non-Abelian QSP sequences.

\textbf{Performance:} We illustrate the circuit of the D/A conversion protocol in Fig.~\ref{fig:Hastrup_state_transfer}b, showing the series of $V_j^\dag W_j^\dag $ operations acting between the oscillator and qubits. We also illustrate in Fig.~\ref{fig_statetransfer_examples_new}b the Wigner function of the final oscillator state after D/A conversion for various initial 3-qubit states. Let us next analyze the gate complexity and performance of this protocol.

As per Eq.~(\ref{eq:Hastrup_state_tr_operator}), this protocol requires $n$ gates, which collectively require a total displacement $ \sum_{j=1}^n \mathcal{O}(\Delta 2^j) = \mathcal{O}(\Delta 2^n)$. As in the previous AD/DA conversion protocol, this implies an overall time complexity $\mathcal{O}(\Delta 2^n)$ when the displacements are implemented with a fixed coupling between the qubits and oscillator, although this can be reduced with a sufficiently tunable and strong coupling.

Next, consider the fidelity of this protocol. Ref.~\cite{hastrup2022universal} presents numerical results on the fidelity achieved in the A/D direction. For example, in transferring the harmonic oscillator eigenstate $|3\rangle$ onto $n$ qubits, the protocol achieves infidelity $\approx 0.2$ for $n=4$, and $\approx 9\cdot 10^{-4}$ for $n=10$, indicating that the performance improves drastically with increasing $n$. Achieving this performance however requires that $\Delta$ be carefully tuned for each value of $n$ to maximize the fidelity, yet no analytical bounds on fidelity are provided in Ref.~\cite{hastrup2022universal} to guide this tuning. Here, we fill this gap by providing fidelity bounds in both the D/A and A/D directions.

In the A/D direction, the exact output state of Eq.~\eqref{eq:HastrupTransferedState} is approximately equal to the desired output state of Eq.~\eqref{eq:HastrupTransferedState_Approx}. The approximations used in reaching this desired state require that $\psi(q)$ have support limited to $|q|\leq \tfrac{\Delta}{2} (2^n-1)$, and be slowly varying as $|\frac{d\psi}{dq}| \ll 1/\Delta$. A careful analysis of this protocol, presented in Sec. SM.III, indicates that the fidelity between these two states, and thus the fidelity of the A/D direction, is
\begin{equation}\label{eq:non-Abelian-fidelity}
\begin{aligned}
     1 &- \mathcal{O} \Bigg( \int_{-\infty}^{-\frac{\Delta}{2}(2^n-1)} dq |\psi(q)|^2 + \int_{\frac{\Delta}{2}(2^n-1)}^{\infty} dq |\psi(q)|^2 \Bigg) \\
      &- \mathcal{O}\Bigg( \Delta \int_{-\frac{\Delta}{2}(2^n-1)}^{\frac{\Delta}{2}(2^n-1)} dq \big| \tfrac{d}{dq}|\psi(q)|^2 \big| \Bigg).
\end{aligned}
\end{equation}
Notably, these two contributions to the infidelity arise precisely from the approximations used in simplifying the exact state to the approximate state. The first contribution depends on the support of $\psi(q)$ outside of $|q|\leq \tfrac{\Delta}{2} (2^n-1)$, and the second on the derivative of $\psi(q)$.

Moreover, D/A conversion is defined by Eq.~\eqref{eq:Hastrup_inverse_state_transfer}. By an analysis similar to the A/D direction, also presented in Sec. SM.III, we find that the fidelity of D/A conversion is
\begin{equation}\label{eq:non_Abelian_ST_Fidelity}
    1- \mathcal{O}\Bigg( \int_{-\infty}^{-\frac{\Delta}{2}(2^n-1)} dq |\psi(q)|^2 + \int_{\frac{\Delta}{2}(2^n-1)}^{\infty} dq |\psi(q)|^2 \Bigg),
\end{equation}
where now $\psi(q) = \frac{1}{\sqrt{\Delta}} \sum_{\mathbf{s}} c_{\mathbf{s}} \text{sinc} \left( \tfrac{\pi (q-q_{\mathbf{s}})}{\Delta} \right)$ is the CV wave function upon ideal D/A conversion. Evidently, in this direction, the fidelity is impeded only by the support of $\psi(q)$ outside $|q| < \tfrac{\Delta}{2} (2^n-1)$. A term analogous to the second contribution in Eq.~\eqref{eq:non-Abelian-fidelity} is absent, because the approximation that produces this contribution is naturally satisfied in the D/A direction; see Sec. SM.III for details. 

In summary, the performance of AD/DA conversion via non-Abelian QSP is encapsulated in the following theorem:
\begin{restatable}[Quantum AD/DA Conversion with Non-Abelian QSP]{theorem}{si} \label{thm:ST_non-Abelian_QSP}
    The quantum D/A conversion protocol based on hybrid non-Abelian QSP achieves a fidelity 
    \begin{equation}
        1 - \mathcal{O} \left( \int_{-\infty}^{-\frac{\Delta}{2}(2^n-1)} dq |\psi(q)|^2 + \int_{\frac{\Delta}{2}(2^n-1)}^{\infty} dq |\psi(q)|^2 \right) ,
    \end{equation}
    at a gate complexity $\mathcal{O}(n)$ and time complexity $\mathcal{O}(\Delta 2^n)$, where $n$ is the number of qubits, $\psi(q) = \frac{1}{\sqrt{\Delta}} \sum_{\mathbf{s}} c_{\mathbf{s}} \mathrm{sinc} \left( \tfrac{\pi (q-q_{\mathbf{s}})}{\Delta} \right)$ is the resulting CV wave function, $\{ c_{\mathbf{s}} \}$ are the coefficients of the initial qubit state $\ket{\psi}_Q = \sum_x c_{\mathbf{s}} |\phi_{\mathbf{s}} \rangle_Q $ being transferred, and $\Delta$ is a spacing parameter.

    Analogously, in reverse this furnishes a quantum A/D conversion protocol with identical gate and time complexity, and a fidelity given by Eq.~\eqref{eq:non-Abelian-fidelity}.
\end{restatable}

Again, for constant $\Delta$, the runtime scales linearly in the dimension of the DV Hilbert space, as anticipated, and owing itself to the encoding of the initial state in $2^n$ basis states equally spaced in position space. Moreover, the fidelity is maximized when the contributions from the above integrals are small. This equates to $\psi(q)$ having limited support over just $|q| \leq \tfrac{\Delta}{2} (2^n-1)$, and being slowly varying as $|\frac{d\psi}{dq}| \ll 1/\Delta$, which are precisely the approximations used in Ref.~\cite{hastrup2022universal} to simplify their results. As such, maximizing the fidelity requires tuning $\Delta$ and $n$ to optimize the above analytical bounds. 

In summary, we have presented two quantum AD/DA protocols that allow sampling of CV wave functions into DV qubits as well as interpolation of DV qubit data into CV wave functions on quantum harmonic oscillators. As discussed in Fig. \ref{fig:CV-DV-schematic}, our results can be understood from a classical signal processing perspective by drawing an analogy between time-frequency duality versus quantum position-momentum duality. For example, the width of the CV wave function in momentum representation $\Phi(p)$ [Fig. \ref{fig:CV-DV-schematic}(b)] serves as a rough `frequency' span of a time-dependent classical signal. While the spacing $\Delta$ between adjacent CV Gaussian or sinc states is analogous to the time interval for sampling classical time-dependent signals. Due to the limited sampling `frequency' ($\sim \frac{1}{\Delta}$), the DV representation of aperiodic CV data will necessarily be periodic with a limited `bandwidth'.

\section{Quantum Fourier Transform from Oscillator Evolution}
\label{sec:qft}
\noindent 

The above quantum AD/DA conversion protocols can be used to implement quantum algorithms on hybrid CV-DV processors. We demonstrate this by using these protocols to implement the quantum Fourier transform (QFT) on CV-DV hardware. The QFT is an important quantum subroutine, ubiquitous in many quantum algorithms such as Shor's algorithm and phase estimation~\cite{Nielsen_Chuang}. It is defined on an $n$-qubit state $|\psi \rangle_Q = \sum_{\mathbf{x}} c_{\mathbf{x}} |\mathbf{x}\rangle_Q $ as the unitary transformation 
\begin{equation}\label{eq:QFT_equation}
    U_{\text{QFT}} |\psi \rangle_Q = \sum_{\mathbf{x}} \Bigg[ \sum_{\mathbf{y}} \frac{1}{\sqrt{2^n}}  c_{\mathbf{y}} e^{2\pi i xy/2^n} \Bigg] |\mathbf{x}\rangle_Q ,
\end{equation} 
which effectively implements a discrete Fourier transform of the coefficients $c_{\mathbf{x}}$. While the traditional construction of the QFT as a DV quantum circuit is well-known~\cite{Nielsen_Chuang}, the construction on CV-DV hardware will differ significantly due to the fundamental differences between oscillators and qubits.

To motivate our construction of the QFT on a CV-DV system, recall that the free evolution of an oscillator swaps position and momentum (see Eq.~\eqref{eq:Fourier_gate2}), thus applying a continuous Fourier transform to the wave function. Using this intuition, we show how QFT can be realized by transferring an initial DV state to a CV state, enacting a free evolution, and finally transferring the state back to the DV system. Importantly however, modifications are required to connect this continuous Fourier transform to the discrete Fourier transform necessitated by the QFT.

Prior work in this direction includes Ref.~\cite{chen2019quantum}, which uses Kerr non-linearities between two oscillators to perform the QFT. However, they encode the qubit states into Fock states on the oscillators, which requires a time an order of magnitude greater than a single photon coherence time, and hence limits their utility. Their algorithm also requires that one perform a photon-number resolved measurement and post-select to disentangle the two oscillators, which requires significant runtime and control. On the other hand, the QFT algorithms we put forth here are not inhibited by these challenges.

In this section, we first describe a correspondence between the continuous Fourier transform and the discrete Fourier transform, which will allow us to connect oscillator evolution and the QFT. We then use this correspondence to develop two algorithms for realizing the QFT on CV-DV hardware incorporating the above AD/DA conversion protocols, and defer the full details to the Supplemental Material Sec. SM.IV.

\subsection{Continuous-Discrete Fourier Transform Correspondence}
\label{ssec:C-D_FT_Correspondence}

Crucial to our construction of the QFT is a correspondence between the continuous Fourier transform and the discrete Fourier transform. Specifically, we show that the continuous Fourier transform of a discrete, periodic signal can reproduce the discrete Fourier transform of the signal. 

To see this, consider a discrete signal $c_x$ for $x \in [0, ..., N-1]$, that is made periodic over $x\in \mathbb{Z}$ and encoded in a continuous function $f(q)$ as
\begin{equation}\label{eq:wave_func_post_transfer}
    f(q) = \sum_{x \in \mathbb{Z}} c_x g(q-x\Delta),
\end{equation}
where $g(q)$ is a basis function localized about $q=0$ and $\Delta$ is the spacing between basis functions. The continuous Fourier transform of this function evaluates to
\begin{equation}
    \tilde{f}(p) = \sum_{x \in \mathbb{Z}} c_x \tilde{g}(p) e^{ipx\Delta},
\end{equation}
where $\tilde{g}(p)$ is the Fourier transform of $g(q)$. We can then split the index $x$ into $x = N k+y$ for $k \in \mathbb{Z}$ and $y \in [0,1,...,N-1]$:
\begin{equation}\label{eq:FT_correspondence}
\begin{aligned}
    &\sum_{k\in \mathbb{Z}} e^{ipNk \Delta}  \sum_{y=0}^{N-1} c_y e^{ipy \Delta} \tilde{g}(p) \\
    &=\sum_{l \in \mathbb{Z}} \tilde{g}(\tfrac{2\pi }{ \Delta} \tfrac{l}{N}) \frac{2 \pi}{N \Delta} \delta(p - \tfrac{2\pi }{ \Delta} \tfrac{l}{N}) \cdot\sum_{y=0}^{N-1}  c_y e^{i 2\pi y l /N} \\
    &=\sum_{l \in \mathbb{Z}} \frac{2 \pi}{\Delta} \tilde{g}(\tfrac{2\pi }{ \Delta} \tfrac{l}{N}) \delta(p - \tfrac{2\pi }{ \Delta} \tfrac{l}{N}) \cdot \tilde{c}_l
\end{aligned}
\end{equation}
where we have noted that $\sum_{k\in \mathbb{Z}} e^{ipNk \Delta} = \frac{2 \pi}{N \Delta} \sum_{l\in \mathbb{Z}} \delta(p - \tfrac{2\pi }{ \Delta}\tfrac{l}{N})$ is a Dirac comb, and we have denoted by $\tilde{c}_l$ the discrete Fourier transform of $c_y$. Evidently then, the continuous Fourier transform of a discrete signal that is encoded periodically in continuous basis functions results in a sum over the discrete Fourier transform of the signal, with coefficients proportional to the Fourier transform of the basis function.

This correspondence can be used to perform the quantum Fourier transform by letting $c_x$ be the coefficients of an initial state on qubits. Then, $f(q)$ represents the wave function of an oscillator after transferring the qubits state with basis function $g(q)$. By enacting a continuous Fourier transform on the oscillator (i.e. free evolution), the new wave function given by Eq.~\eqref{eq:FT_correspondence} will pluck out the states $|q = \tfrac{2\pi l}{N \Delta} \rangle$ with coefficients proportional to the discrete Fourier transform of $c_x$. As this discrete Fourier transform equates to the coefficients of QFT, we find that appropriately transferring this state back to qubits produces the QFT of the initial state. 

\subsection{Quantum Fourier Transform Protocols}
\label{ssec:QFT}

We now use the above correspondence and intuition to present algorithms for the QFT. The first step is to make the quantum state periodic, such that we can invoke the correspondence. We can achieve this by prepending the initial state with $a$ ancilla qubits $|+\rangle^{\otimes a}$. To see this, consider an initial $n$-qubit state $|\psi \rangle_Q = \sum_{\mathbf{x}} c_{\mathbf{x}} |{\mathbf{x}}\rangle_Q $, and prepend it with $|+\rangle^{\otimes a}_Q$:
\begin{equation}\label{eq:QSPTransfer}
\begin{aligned}
    \left(|+\rangle^{\otimes a} |\psi\rangle \right)_Q  & = \frac{1}{\sqrt{2^a}}\sum_{k=0}^{2^a-1}\sum_{x=0}^{2^n-1} c_{\mathbf{x}} |2^n\cdot k + x \rangle_Q .
\end{aligned}
\end{equation}
The coefficients are now $c_{\mathbf{x}}/\sqrt{2^a}$ and repeat over $2^a$ periods of size $2^n$. This renders the coefficients (approximately) periodic and enables use of the above correspondence. As we will see, increasing the number of ancilla qubits makes the state more periodic and will increase the fidelity with the exact QFT.

\begin{figure}[htbp]
    \begin{center}
    \includegraphics[width=0.99\linewidth]{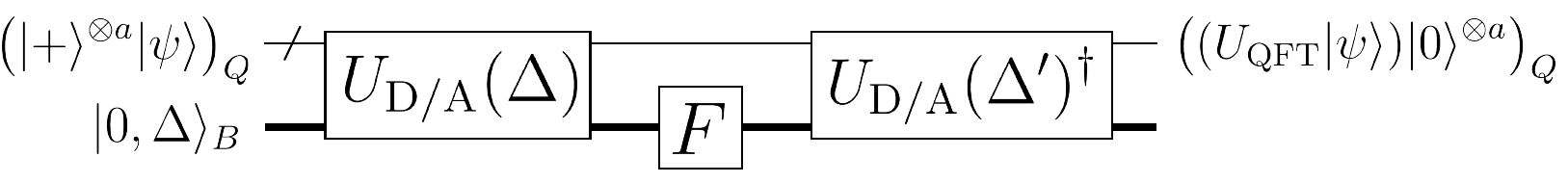}
    \end{center}
    \caption{The generic circuit used to implement the quantum Fourier transform of an $n$-qubit state $|\psi\rangle$ using AD/DA conversion. The initial state is first appended with ancilla qubits $|+\rangle^{\otimes a}$ to facilitate the QFT, as discussed in the main text. The unitaries $U_{\text{D/A}}(\Delta)$ are A/D and D/A conversion unitaries respectively (either single variable or non-Abelian), and $F$ is the Fourier gate of Eq.~\eqref{eq:Fourier_gate}. The circuit outputs the state $(U_{\text{QFT}}|\psi\rangle) |0\rangle^{\otimes a}$, from which the QFT may be obtained. For precise details on the construction of the QFT, see Supplemental Material Sec.~SM.IV.} 
    \label{fig:QFT_Generic}
\end{figure}

The remaining steps of the QFT algorithm are as suggested in the preamble of this section. We use a D/A conversion protocol (either single-variable or non-Abelian QSP) to transfer the state $\left(|+\rangle^{\otimes a} |\psi\rangle \right)_Q $ to an oscillator state with spacing $\Delta$, and then apply the Fourier gate to the oscillator (equivalent to free evolution, see Eq.~\eqref{eq:Fourier_gate2}). Finally, we use the corresponding A/D conversion to transfer the oscillator state back to qubits with a reciprocal spacing $\sim \frac{\pi}{2^n \Delta}$. This reciprocal spacing is chosen to match the behavior of the original spacing upon Fourier transform, and correctly implements the QFT. For intuitive purposes, we illustrate the corresponding circuit in Fig.~\ref{fig:QFT_Generic}.

Both QFT protocols we introduce here follow this structure, but incorporate modifications to accommodate their respective AD/DA conversion, such as additional gates and ancilla qubits. We present the full detailed protocols in Supplemental Material Sec. SM.IV., and prove the following two theorems on their performance. First, using the single-variable QSP AD/DA conversion protocol to perform the QFT, we have:
\begin{restatable}[QFT from Oscillator Evolution and AD/DA Conversion (Single-Variable QSP]{theorem}{qft_single_var}
    \label{thm:qft_single_var}
    Using the single-variable QSP AD/DA conversion protocol, oscillator evolution, and additional bosonic gates one can realize the quantum Fourier transform of an $n$ qubit state with fidelity $1 - \mathcal{O}((n+a)\epsilon) - \mathcal{O}(e^{-\mathcal{O}(\Delta^2/\sigma^2)} ) - \mathcal{O}(1/2^a)$, gate complexity of $\mathcal{O}( 2^{n+a}\log(1/\epsilon))$, and time complexity $\mathcal{O}(2^{n+a}(\Delta+\log(1/\epsilon)))$, where $a$ is the number of ancilla qubits, $\epsilon$ is the error of the QSP polynomial, and $\Delta$ is a spacing parameter. 
\end{restatable}
Evidently, the fidelity is maximized in the limit of small QSP error $\epsilon$, small relative spacing $\sigma/\Delta$, and many ancilla qubits. However, this choice of parameters will also increase the gate and time complexity as explained above.

Similarly, using the non-Abelian QSP AD/DA conversion protocol to perform the QFT, we have:
\begin{restatable}[QFT from Oscillator Evolution and AD/DA Conversion Via Non-Abelian QSP]{theorem}{qft_non_Abelian}
    \label{thm:qft_non_Abelian}
    Using AD/DA conversion via hybrid non-Abelian QSP , oscillator evolution, and additional bosonic gates one can realize the quantum Fourier transform of an $n$ qubit state with fidelity $1 - \mathcal{O}(1/2^a)$, gate complexity $\mathcal{O}(n+a)$, and time complexity $\mathcal{O}(\Delta 2^{n+a})$, where $a$ is the number of ancilla qubits and $\Delta$ is a spacing parameter.
\end{restatable}
Here the fidelity is maximized in the limit of many ancilla qubits, which also increases the gate and time complexity.

Lastly, we note that these protocols can also be used to implement the inverse QFT, which acts as $U_{\text{QFT}}^\dag |\textbf{y} \rangle_Q = \frac{1}{\sqrt{2^n}} \sum_{\mathbf{x}} e^{-2\pi i xy/2^n}  |\mathbf{x}\rangle_Q $. Because the inverse QFT differs from the ordinary QFT by only the sign of the phases, it can be implemented through our protocol by instead performing an inverse free evolution $F^\dag$ on the CV mode. This corresponds to a phase space rotation by $-\pi/2$, and introduces the appropriate phases necessary to reconstruct the inverse QFT. The associated error is equivalent to the infidelity incurred in the forward QFT implementation.

\subsection{Possible Experimental Realizations}

The quantum gates used in our protocol include single-qubit rotations (Eq. \eqref{eq:su2}), conditional displacement gate (Eqs. \eqref{wz_k} and \eqref{wz_lambda}), free-evolution gate (Eq. \eqref{eq:Fourier_gate}) of oscillators. All these gates can be realized on leading quantum hardware, including trapped-ion \cite{bruzewicz2019trapped} and superconducting platforms \cite{BlaiscQEDReviewRMP2020}. For example, with trapped ions, the ions' vibrational motion encodes the quantum harmonic oscillators, while its internal energy levels (hyperfine or electronic states) represent the qubits~\cite{bruzewicz2019trapped}. Conditional displacements can be realized with a bichromatic field which simultaneously drives red- and blue-sideband operations on the ion. Single-qubit rotations can be realized by applying a laser, microwave or RF drive, depending on the internal energy levels of the ions (dipole or magnetic transitions) used to encode the qubits. In a superconducting platform, the quantum oscillators are represented by microwave photons in a high-fidelity aluminum cavity, while the qubits can be realized by transmons \cite{BlaiscQEDReviewRMP2020}. The conditional displacement gate can be engineered from dispersive coupling between the oscillator and qubit, as demonstrated in Ref. \cite{EickbuschECD}. In both cases, the oscillator free-evolution is a native gate, and can be switched on and off by switching between the lab reference frame and the oscillator free-evolution reference frame. 

In practice, physical realizations of quantum oscillators and qubits are noisy. This puts an upper bound on the depth of our hybrid CV-DV circuits that can be executed on real hardware. The effects of noise are platform-specific. In a superconducting platform, the lifetime of qubits ($100-500$ $\mu$s) is usually substantially shorter than that of cavities ($10^4-10^6$ $\mu$s) \cite{PRXQuantum.4.030336,liu2024hybrid}. In a trapped-ion platform, the lifetime of the ions' motion is typically on the order of 10 $m$s \cite{simon2020measuring}, limited by heating from the trapping potential. The qubit lifetime on trapped ions, on the other hand, can be extremely long ($>$ 1 hour \cite{wang2021single}). A quantitative analysis of the effect of noise on our protocol requires a detailed circuit-level simulation, which we leave for future work. Nevertheless, using a trapped-ion platform with $\sim10$ ions and $\sim 10$ modes, it is possible to perform a proof-of-principle demonstration of our protocol on real quantum hardware in the near future. However, executing our protocol with a very deep circuit and high fidelity will likely require quantum error corrected qubits \cite{devitt2013quantum} and robust oscillators \cite{noh2020encoding}. 

\section{Conclusion and Outlook}
\label{sec:conclusion-outlook}

In this paper, we have established a framework of mixed analog-digital QSP for execution on hybrid CV-DV quantum hardware. These algorithms generate polynomial transformations of position and momentum, and open the door to a wide variety of algorithms on CV-DV quantum processors. We used this framework to present two unitary protocols that convert a DV quantum state to a CV quantum state and vice-versa (Theorems~\ref{thm:ST_single_var_QSP} and ~\ref{thm:ST_non-Abelian_QSP}), thus furnishing a quantum counterpart to AD/DA conversion in classical signal processing. We established the gate and time complexity of both AD/DA conversion protocols; notably, the protocol based on hybrid non-Abelian QSP achieves an efficient gate count of $O(n)$ for converting between an $n$-qubit state and a CV state. 

As a further contribution, we demonstrated how this framework can realize the quantum Fourier transform of an $n$-qubit state by simply transferring the qubits state to an oscillator, letting the oscillator undergo free-evolution, and then transferring the state back to the qubits (Theorems~\ref{thm:qft_single_var} and~\ref{thm:qft_non_Abelian}). Importantly, the protocol incorporating non-Abelian QSP requires only $O(n)$ hybrid CV-DV gates to implement the QFT, as opposed to the $O(n^2)$ gates of the conventional construction of the QFT~\cite{Nielsen_Chuang}.

Despite these results, ample open questions remain to be addressed. First, while we have shown that non-Abelian QSP offers more efficient protocols for AD/DA conversion and the QFT compared to single-variable QSP, a complete theory of non-Abelian QSP remains to be established, fundamentally hinging upon an extension of QSP to the multivariate setting. 
Second, while classical digital signal processing benefits from its robustness, it is not clear how to make analog-digital QSP robust against the noise afflicting quantum oscillators and qubits. One idea is to generalize the connection between classical signal processing, frames and wavelet theory~\cite{christensen2003introduction} to the quantum setting. Conversely, frames and wavelets could also guide the design of novel QSP algorithms. Moreover, state-of-the-art advanced sampling techniques in classical signal processing such as non-uniform sampling and compressed sensing \cite{aldroubi2001nonuniform,marvasti2012unified,4472240,grochenig2001foundations,1003065,baron2009bayesian} are unknown in mixed analog-digital quantum signal processing. Evaluating the possibilities of generalizing these non-uniform sampling strategies to sample CV wave functions to DV qubits in the quantum case and developing efficient quantum circuits to achieve so will be another milestone for mixed analog-digital QSP. Once realized, this will dramatically increase the efficiency of processing quantum CV signals using DV quantum resources. 

In addition, while our implementation of the QFT with non-Abelian QSP achieves a gate count linear in the number of qubits, there exist more gate-efficient constructions of the QFT on qubits that use parallelization~\cite{cleve2000fast}. Accordingly, it remains an open question how to parallelize our QFT construction over multiple oscillators, which could perhaps prove advantageous. In addition, it would be interesting to investigate what other algorithms and operations could be realized through analog-digital QSP. For instance, analogous to our QFT constructions, the fractional quantum Fourier transform \cite{weimann2016implementation} could be realized by letting the oscillator evolve for only a fraction of its period. Likewise, given the unification of quantum algorithms afforded by QSP, it appears promising that analog-digital QSP could act as a Rosetta Stone for translating quantum algorithms from DV hardware to hybrid CV-DV hardware.

Just as the fundamental roles that analog and digital signal processing play in classical computing, our framework provides a concrete way to process mixed analog-digital quantum signals and opens a new direction for signal processing relevant applications. One example of such analog quantum data is electromagnetic waves that may arise in quantum radar \cite{assouly2023quantum,lanzagorta2011quantum} and wireless communications \cite{heath2016overview}. Despite the fact that current antennas are mostly classical, advancements in quantum hardware have opened opportunities to explore quantum effects of these EM waves with quantum antennas \cite{mikki2020quantum,gurses2024free}. 
There are also opportunities to design quantum matched filters and filter banks to enhance detection and estimation sensitivity \cite{PhysRevResearch.4.023006,9951210}. Although we focus on static time-independent quantum signals, generalizing our framework to process time-dependent quantum signals is another interesting direction \cite{10887919,10890646}. More broadly, 
it would be exciting to synthesize our QSP framework with classical mixed signal processing \cite{kester2003mixed} to develop a theory of quantum-classical signal processing. Beyond theoretical work, co-designing these novel signal processing frameworks with hardware and chips could harness quantum information to make real-world impacts in the forthcoming quantum era of electrical and computer engineering.

\section*{Acknowledgment}

The authors thank Nathan Wiebe and Shraddha Singh for helpful discussions. YL, ILC, and SMG acknowledge the support by the U.S. Department of Energy, Office of Science, National Quantum Information Science Research Centers, Co-design Center for Quantum Advantage (C$^2$QA) under contract number DE-SC0012704.
SMG acknowledges support by the Army Research Office (ARO) under Grant Number W911NF-23-1-0051. JS was supported in part by the Army Research Office under the CVQC project W911NF-17-1-0481 and NTT Research. YL was supported in part by NTT Research and the U.S. Department of Energy, Office of Science, Office of Advanced Scientific Computing Research (ASCR), under Award Number DE-SC0025384.

External interest disclosure: SMG is a consultant for, and an equity holder in, Quantum Circuits, Inc.

\bibliographystyle{IEEEtran}
\bibliography{IEEEabrv,ref}

\begin{thebibliography}{10}
\providecommand{\url}[1]{#1}
\csname url@samestyle\endcsname
\providecommand{\newblock}{\relax}
\providecommand{\bibinfo}[2]{#2}
\providecommand{\BIBentrySTDinterwordspacing}{\spaceskip=0pt\relax}
\providecommand{\BIBentryALTinterwordstretchfactor}{4}
\providecommand{\BIBentryALTinterwordspacing}{\spaceskip=\fontdimen2\font plus
\BIBentryALTinterwordstretchfactor\fontdimen3\font minus \fontdimen4\font\relax}
\providecommand{\BIBforeignlanguage}[2]{{%
\expandafter\ifx\csname l@#1\endcsname\relax
\typeout{** WARNING: IEEEtran.bst: No hyphenation pattern has been}%
\typeout{** loaded for the language `#1'. Using the pattern for}%
\typeout{** the default language instead.}%
\else
\language=\csname l@#1\endcsname
\fi
#2}}
\providecommand{\BIBdecl}{\relax}
\BIBdecl

\bibitem{proakis2007digital}
J.~G. Proakis, \emph{Digital signal processing: principles, algorithms, and applications, 4/E}.\hskip 1em plus 0.5em minus 0.4em\relax Pearson Education India, 2007.

\bibitem{oppenheim1999discrete}
A.~V. Oppenheim, \emph{Discrete-time signal processing}.\hskip 1em plus 0.5em minus 0.4em\relax Pearson Education India, 1999.

\bibitem{van1992computational}
C.~Van~Loan, \emph{Computational frameworks for the fast Fourier transform}.\hskip 1em plus 0.5em minus 0.4em\relax SIAM, 1992.

\bibitem{marks2012introduction}
R.~J.~I. Marks, \emph{Introduction to Shannon sampling and interpolation theory}.\hskip 1em plus 0.5em minus 0.4em\relax Springer Science \& Business Media, 2012.

\bibitem{winder2002analog}
S.~Winder, \emph{Analog and digital filter design}.\hskip 1em plus 0.5em minus 0.4em\relax Elsevier, 2002.

\bibitem{haensch2018next}
\BIBentryALTinterwordspacing
W.~Haensch, T.~Gokmen, and R.~Puri, ``The next generation of deep learning hardware: Analog computing,'' \emph{Proceedings of the IEEE}, vol. 107, no.~1, pp. 108--122, 2018. [Online]. Available: \url{https://doi.org/10.1109/JPROC.2018.2871057}
\BIBentrySTDinterwordspacing

\bibitem{gold2011speech}
B.~Gold, N.~Morgan, and D.~Ellis, \emph{Speech and audio signal processing: processing and perception of speech and music}.\hskip 1em plus 0.5em minus 0.4em\relax John Wiley \& Sons, 2011.

\bibitem{kester2003mixed}
W.~Kester, \emph{Mixed-signal and DSP design techniques}.\hskip 1em plus 0.5em minus 0.4em\relax Newnes, 2003.

\bibitem{sharpeshkar2010ultra}
R.~Sharpeshkar, \emph{Ultra low power bioelectronics: Fundamentals, biomedical applications, and bio-inspired system}.\hskip 1em plus 0.5em minus 0.4em\relax Cambridge University Press, 2010.

\bibitem{guo2016energy}
\BIBentryALTinterwordspacing
N.~Guo, Y.~Huang, T.~Mai, S.~Patil, C.~Cao, M.~Seok, S.~Sethumadhavan, and Y.~Tsividis, ``Energy-efficient hybrid analog/digital approximate computation in continuous time,'' \emph{IEEE Journal of Solid-State Circuits}, vol.~51, no.~7, pp. 1514--1524, 2016. [Online]. Available: \url{https://doi.org/10.1109/JSSC.2016.2543729}
\BIBentrySTDinterwordspacing

\bibitem{Nielsen_Chuang}
M.~A. Nielsen and I.~L. Chuang, \emph{Quantum Computation and Quantum Information}.\hskip 1em plus 0.5em minus 0.4em\relax {Cambridge University Press}, 2010.

\bibitem{eldar2002quantum}
\BIBentryALTinterwordspacing
Y.~C. Eldar and A.~V. Oppenheim, ``Quantum signal processing,'' \emph{IEEE Signal Processing Magazine}, vol.~19, no.~6, pp. 12--32, 2002. [Online]. Available: \url{https://doi.org/10.1109/MSP.2002.1043298}
\BIBentrySTDinterwordspacing

\bibitem{low2016methodology}
\BIBentryALTinterwordspacing
G.~H. Low, T.~J. Yoder, and I.~L. Chuang, ``Methodology of resonant equiangular composite quantum gates,'' \emph{Phys. Rev. X}, vol.~6, p. 041067, Dec 2016. [Online]. Available: \url{https://link.aps.org/doi/10.1103/PhysRevX.6.041067}
\BIBentrySTDinterwordspacing

\bibitem{low2017optimal}
\BIBentryALTinterwordspacing
G.~H. Low and I.~Chuang, ``Optimal hamiltonian simulation by quantum signal processing,'' \emph{Phys. Rev. Lett.}, vol. 118, p. 010501, 2017. [Online]. Available: \url{https://link.aps.org/doi/10.1103/PhysRevLett.118.010501}
\BIBentrySTDinterwordspacing

\bibitem{gilyen2019quantum}
\BIBentryALTinterwordspacing
A.~Gily{\'e}n, Y.~Su, G.~H. Low, and N.~Wiebe, ``Quantum singular value transformation and beyond: exponential improvements for quantum matrix arithmetics,'' in \emph{Proceedings of the 51st Annual ACM SIGACT Symposium on Theory of Computing}, 2019, pp. 193--204. [Online]. Available: \url{https://doi.org/10.1145/3313276.3316366}
\BIBentrySTDinterwordspacing

\bibitem{GrandUnificationAlgos}
\BIBentryALTinterwordspacing
J.~M. Martyn, Z.~M. Rossi, A.~K. Tan, and I.~L. Chuang, ``Grand unification of quantum algorithms,'' \emph{PRX Quantum}, vol.~2, p. 040203, Dec 2021. [Online]. Available: \url{https://link.aps.org/doi/10.1103/PRXQuantum.2.040203}
\BIBentrySTDinterwordspacing

\bibitem{rossi2021mqsp}
\BIBentryALTinterwordspacing
Z.~M. Rossi and I.~L. Chuang, ``Multivariable quantum signal processing (m-qsp): prophecies of the two-headed oracle,'' \emph{Quantum}, vol.~6, p. 811, 2022. [Online]. Available: \url{https://quantum-journal.org/papers/q-2022-09-20-811/}
\BIBentrySTDinterwordspacing

\bibitem{motlagh2023generalized}
D.~Motlagh and N.~Wiebe, ``Generalized quantum signal processing,'' \emph{arXiv preprint arXiv:2308.01501}, 2023.

\bibitem{rossi2023quantum}
\BIBentryALTinterwordspacing
Z.~M. Rossi, V.~M. Bastidas, W.~J. Munro, and I.~L. Chuang, ``Quantum signal processing with continuous variables,'' 2023. [Online]. Available: \url{https://arxiv.org/abs/2304.14383}
\BIBentrySTDinterwordspacing

\bibitem{tan2023perturbative}
\BIBentryALTinterwordspacing
A.~K. Tan, Y.~Liu, M.~C. Tran, and I.~L. Chuang, ``Perturbative model of noisy quantum signal processing,'' \emph{Physical Review A}, vol. 107, no.~4, p. 042429, 2023. [Online]. Available: \url{https://doi.org/10.1103/PhysRevA.107.042429}
\BIBentrySTDinterwordspacing

\bibitem{chao2020finding}
\BIBentryALTinterwordspacing
R.~Chao, D.~Ding, A.~Gilyen, C.~Huang, and M.~Szegedy, ``Finding angles for quantum signal processing with machine precision,'' 2020. [Online]. Available: \url{https://arxiv.org/abs/2003.02831}
\BIBentrySTDinterwordspacing

\bibitem{WhaleyRobust}
\BIBentryALTinterwordspacing
Y.~Dong, X.~Meng, K.~B. Whaley, and L.~Lin, ``Efficient phase-factor evaluation in quantum signal processing,'' \emph{Phys. Rev. A}, vol. 103, p. 042419, Apr 2021. [Online]. Available: \url{https://link.aps.org/doi/10.1103/PhysRevA.103.042419}
\BIBentrySTDinterwordspacing

\bibitem{ying2022stable}
\BIBentryALTinterwordspacing
L.~Ying, ``Stable factorization for phase factors of quantum signal processing,'' \emph{Quantum}, vol.~6, p. 842, 2022. [Online]. Available: \url{https://doi.org/10.22331/q-2022-10-20-842}
\BIBentrySTDinterwordspacing

\bibitem{martyn2023efficient}
\BIBentryALTinterwordspacing
J.~M. Martyn, Y.~Liu, Z.~E. Chin, and I.~L. Chuang, ``{Efficient fully-coherent quantum signal processing algorithms for real-time dynamics simulation},'' \emph{The Journal of Chemical Physics}, vol. 158, no.~2, p. 024106, 01 2023. [Online]. Available: \url{https://doi.org/10.1063/5.0124385}
\BIBentrySTDinterwordspacing

\bibitem{dong_beyond_2022}
\BIBentryALTinterwordspacing
Y.~Dong, J.~Gross, and M.~Y. Niu, ``Beyond heisenberg limit quantum metrology through quantum signal processing,'' \emph{arXiv preprint arXiv:2209.11207}, 2022. [Online]. Available: \url{https://arxiv.org/abs/2209.11207}
\BIBentrySTDinterwordspacing

\bibitem{10887919}
\BIBentryALTinterwordspacing
A.~Majumdar, B.~N. Bakalov, D.~Baron, and Y.~Liu, ``Implementing finite impulse response filters on quantum computers,'' in \emph{ICASSP 2025 - 2025 IEEE International Conference on Acoustics, Speech and Signal Processing (ICASSP)}, 2025, pp. 1--5. [Online]. Available: \url{https://doi.org/10.1109/ICASSP49660.2025.10887919}
\BIBentrySTDinterwordspacing

\bibitem{10434630}
\BIBentryALTinterwordspacing
A.~Karthik, A.~V, G.~Nijhawan, A.~Rana, I.~Khan, and Z.~L. Naser, ``Quantum signal processing: A new frontier in information processing,'' in \emph{2023 10th IEEE Uttar Pradesh Section International Conference on Electrical, Electronics and Computer Engineering (UPCON)}, vol.~10, 2023, pp. 1527--1532. [Online]. Available: \url{https://doi.org/10.1109/UPCON59197.2023.10434630}
\BIBentrySTDinterwordspacing

\bibitem{10721632}
\BIBentryALTinterwordspacing
A.~M.~A. Sabaawi, M.~R. Almasaoodi, and S.~Imre, ``Advancing quantum communications: Q-ofdm with quantum fourier transforms for enhanced signal integrity,'' in \emph{2024 International Conference on Software, Telecommunications and Computer Networks (SoftCOM)}, 2024, pp. 1--6. [Online]. Available: \url{https://doi.org/10.23919/SoftCOM62040.2024.10721632}
\BIBentrySTDinterwordspacing

\bibitem{10593013}
\BIBentryALTinterwordspacing
S.~Anjimoon, S.~Baswaraju, R.~Sobti, S.~Ajmera, A.~Rana, and A.~A. Hameed, ``Hybrid quantum-classical approaches to optimize signal processing in massive mimo arrays,'' in \emph{2024 International Conference on Communication, Computer Sciences and Engineering (IC3SE)}, 2024, pp. 1--6. [Online]. Available: \url{https://doi.org/10.1109/IC3SE62002.2024.10593013}
\BIBentrySTDinterwordspacing

\bibitem{10670345}
V.~Patel and Z.~Jiang, ``An initial survey on quantum enhanced rf signal extraction in cluttered environments,'' in \emph{NAECON 2024 - IEEE National Aerospace and Electronics Conference}, 2024, pp. 136--141.

\bibitem{shukla2022hybrid}
A.~Shukla and P.~Vedula, ``A hybrid classical-quantum algorithm for digital image processing,'' \emph{Quantum Information Processing}, vol.~22, no.~1, p.~3, 2022.

\bibitem{liu2016power}
\BIBentryALTinterwordspacing
N.~Liu, J.~Thompson, C.~Weedbrook, S.~Lloyd, V.~Vedral, M.~Gu, and K.~Modi, ``Power of one qumode for quantum computation,'' \emph{Phys. Rev. A}, vol.~93, p. 052304, May 2016. [Online]. Available: \url{https://link.aps.org/doi/10.1103/PhysRevA.93.052304}
\BIBentrySTDinterwordspacing

\bibitem{Braunstein2005}
\BIBentryALTinterwordspacing
S.~L. Braunstein and P.~van Loock, ``Quantum information with continuous variables,'' \emph{Rev. Mod. Phys.}, vol.~77, pp. 513--577, Jun 2005. [Online]. Available: \url{https://link.aps.org/doi/10.1103/RevModPhys.77.513}
\BIBentrySTDinterwordspacing

\bibitem{mikki2020quantum}
\BIBentryALTinterwordspacing
S.~Mikki, ``Quantum antenna theory for secure wireless communications,'' in \emph{2020 14th European Conference on Antennas and Propagation (EuCAP)}.\hskip 1em plus 0.5em minus 0.4em\relax IEEE, 2020, pp. 1--4. [Online]. Available: \url{https://doi.org/10.23919/EuCAP48036.2020.9135570}
\BIBentrySTDinterwordspacing

\bibitem{Campagne-Ibarcq2020}
\BIBentryALTinterwordspacing
P.~Campagne-Ibarcq, A.~Eickbusch, S.~Touzard, E.~Zalys-Geller, N.~E. Frattini, V.~V. Sivak, P.~Reinhold, S.~Puri, S.~Shankar, R.~J. Schoelkopf, L.~Frunzio, M.~Mirrahimi, and M.~H. Devoret, ``Quantum error correction of a qubit encoded in grid states of an oscillator,'' \emph{Nature}, vol. 584, no. 7821, pp. 368--372, 2020. [Online]. Available: \url{https://doi.org/10.1038/s41586-020-2603-3}
\BIBentrySTDinterwordspacing

\bibitem{Sivak_GKP_2022}
\BIBentryALTinterwordspacing
V.~V. Sivak, A.~Eickbusch, B.~Royer, S.~Singh, I.~Tsioutsios, S.~Ganjam, A.~Miano, B.~L. Brock, A.~Z. Ding, L.~Frunzio, S.~M. Girvin, R.~J. Schoelkopf, and M.~H. Devoret, ``Real-time quantum error correction beyond break-even,'' 2023. [Online]. Available: \url{https://doi.org/10.1038/s41586-023-05782-6}
\BIBentrySTDinterwordspacing

\bibitem{LuyanSun2020}
\BIBentryALTinterwordspacing
Y.~Ma, Y.~Xu, X.~Mu, W.~Cai, L.~Hu, W.~Wang, X.~Pan, H.~Wang, Y.~P. Song, C.~L. Zou, and L.~Sun, ``Error-transparent operations on a logical qubit protected by quantum error correction,'' \emph{Nature Physics}, 2020. [Online]. Available: \url{https://doi.org/10.1038/s41567-020-0893-x}
\BIBentrySTDinterwordspacing

\bibitem{ni2022beating}
\BIBentryALTinterwordspacing
Z.~Ni, S.~Li, X.~Deng, Y.~Cai, L.~Zhang, W.~Wang, Z.-B. Yang, H.~Yu, F.~Yan, S.~Liu, C.-L. Zou, L.~Sun, S.-B. Zheng, Y.~Xu, and D.~Yu, ``Beating the break-even point with a discrete-variable-encoded logical qubit,'' \emph{Nature}, 2023. [Online]. Available: \url{https://doi.org/10.1038/s41586-023-05784-4}
\BIBentrySTDinterwordspacing

\bibitem{EickbuschECD}
\BIBentryALTinterwordspacing
A.~Eickbusch, V.~Sivak, A.~Z. Ding, S.~S. Elder, S.~R. Jha, J.~Venkatraman, B.~Royer, S.~M. Girvin, R.~J. Schoelkopf, and M.~H. Devoret, ``Fast universal control of an oscillator with weak dispersive coupling to a qubit,'' \emph{Nature Physics}, vol.~18, no.~12, pp. 1464--1469, 2022. [Online]. Available: \url{https://doi.org/10.1038/s41567-022-01776-9}
\BIBentrySTDinterwordspacing

\bibitem{Weedbrook2012}
\BIBentryALTinterwordspacing
C.~Weedbrook, S.~Pirandola, R.~Garc\'{\i}a-Patr\'on, N.~J. Cerf, T.~C. Ralph, J.~H. Shapiro, and S.~Lloyd, ``Gaussian quantum information,'' \emph{Rev. Mod. Phys.}, vol.~84, pp. 621--669, May 2012. [Online]. Available: \url{http://link.aps.org/doi/10.1103/RevModPhys.84.621}
\BIBentrySTDinterwordspacing

\bibitem{andersen2015hybrid}
\BIBentryALTinterwordspacing
U.~L. Andersen, J.~S. Neergaard-Nielsen, P.~Van~Loock, and A.~Furusawa, ``Hybrid discrete-and continuous-variable quantum information,'' \emph{Nature Physics}, vol.~11, no.~9, pp. 713--719, 2015. [Online]. Available: \url{http://dx.doi.org/10.1038/nphys3410}
\BIBentrySTDinterwordspacing

\bibitem{Krastanov2015}
\BIBentryALTinterwordspacing
S.~Krastanov, V.~V. Albert, C.~Shen, C.-L. Zou, R.~W. Heeres, B.~Vlastakis, R.~J. Schoelkopf, and L.~Jiang, ``Universal control of an oscillator with dispersive coupling to a qubit,'' \emph{Phys. Rev. A}, vol.~92, p. 040303, Oct 2015. [Online]. Available: \url{http://link.aps.org/doi/10.1103/PhysRevA.92.040303}
\BIBentrySTDinterwordspacing

\bibitem{QuditsfromOscillatorsPhysRevA.104.032605}
\BIBentryALTinterwordspacing
Y.~Liu, J.~Sinanan-Singh, M.~T. Kearney, G.~Mintzer, and I.~L. Chuang, ``Constructing qudits from infinite-dimensional oscillators by coupling to qubits,'' \emph{Phys. Rev. A}, vol. 104, p. 032605, Sep 2021. [Online]. Available: \url{https://link.aps.org/doi/10.1103/PhysRevA.104.032605}
\BIBentrySTDinterwordspacing

\bibitem{liu2024hybrid}
\BIBentryALTinterwordspacing
Y.~Liu, S.~Singh, K.~C. Smith, E.~Crane, J.~M. Martyn, A.~Eickbusch, A.~Schuckert, R.~D. Li, J.~Sinanan-Singh, M.~B. Soley, T.~Tsunoda, I.~L. Chuang, N.~Wiebe, and S.~M. Girvin, ``Hybrid oscillator-qubit quantum processors: Instruction set architectures, abstract machine models, and applications,'' \emph{arXiv:2407.10381}, 2024. [Online]. Available: \url{https://arxiv.org/abs/2407.10381}
\BIBentrySTDinterwordspacing

\bibitem{C2QA-LGTpaper}
\BIBentryALTinterwordspacing
E.~Crane, K.~C. Smith, T.~Tomesh, A.~Eickbusch, J.~M. Martyn, S.~Kühn, L.~Funcke, M.~A. DeMarco, I.~L. Chuang, N.~Wiebe, A.~Schuckert, and S.~M. Girvin, ``Hybrid oscillator-qubit quantum processors: Simulating fermions, bosons, and gauge fields,'' 2024. [Online]. Available: \url{https://arxiv.org/abs/2409.03747}
\BIBentrySTDinterwordspacing

\bibitem{qspi2023}
\BIBentryALTinterwordspacing
J.~Sinanan-Singh, G.~L. Mintzer, I.~L. Chuang, and Y.~Liu, ``Single-shot quantum signal processing interferometry,'' \emph{arXiv preprint arXiv:2311.13703}, 2023. [Online]. Available: \url{https://arxiv.org/abs/2311.13703}
\BIBentrySTDinterwordspacing

\bibitem{singh2025nonabelian}
\BIBentryALTinterwordspacing
S.~Singh, B.~Royer, and S.~M. Girvin, ``Towards non-abelian quantum signal processing: Efficient control of hybrid continuous- and discrete-variable architectures,'' 2025. [Online]. Available: \url{https://arxiv.org/abs/2504.19992}
\BIBentrySTDinterwordspacing

\bibitem{marvasti2012unified}
\BIBentryALTinterwordspacing
F.~Marvasti, A.~Amini, F.~Haddadi, M.~Soltanolkotabi, B.~H. Khalaj, A.~Aldroubi, S.~Sanei, and J.~Chambers, ``A unified approach to sparse signal processing,'' \emph{EURASIP journal on advances in signal processing}, vol. 2012, pp. 1--45, 2012. [Online]. Available: \url{https://doi.org/10.1186/1687-6180-2012-44}
\BIBentrySTDinterwordspacing

\bibitem{4472240}
\BIBentryALTinterwordspacing
E.~J. Candes and M.~B. Wakin, ``An introduction to compressive sampling,'' \emph{IEEE Signal Processing Magazine}, vol.~25, no.~2, pp. 21--30, 2008. [Online]. Available: \url{https://doi.org/10.1109/MSP.2007.914731}
\BIBentrySTDinterwordspacing

\bibitem{grochenig2001foundations}
K.~Gr{\"o}chenig, \emph{Foundations of time-frequency analysis}.\hskip 1em plus 0.5em minus 0.4em\relax Springer Science \& Business Media, 2001.

\bibitem{1003065}
\BIBentryALTinterwordspacing
M.~Vetterli, P.~Marziliano, and T.~Blu, ``Sampling signals with finite rate of innovation,'' \emph{IEEE Transactions on Signal Processing}, vol.~50, no.~6, pp. 1417--1428, 2002. [Online]. Available: \url{https://doi.org/10.1109/TSP.2002.1003065}
\BIBentrySTDinterwordspacing

\bibitem{kitaev2008wavefunction}
\BIBentryALTinterwordspacing
A.~Kitaev and W.~A. Webb, ``Wavefunction preparation and resampling using a quantum computer,'' 2009. [Online]. Available: \url{https://arxiv.org/abs/0801.0342}
\BIBentrySTDinterwordspacing

\bibitem{hastrup2021improved}
\BIBentryALTinterwordspacing
J.~Hastrup and U.~L. Andersen, ``Improved readout of qubit-coupled gottesman–kitaev–preskill states,'' \emph{Quantum Science and Technology}, vol.~6, no.~3, p. 035016, Jun. 2021. [Online]. Available: \url{http://dx.doi.org/10.1088/2058-9565/ac070d}
\BIBentrySTDinterwordspacing

\bibitem{aldroubi2001nonuniform}
\BIBentryALTinterwordspacing
A.~Aldroubi and K.~Gr{\"o}chenig, ``Nonuniform sampling and reconstruction in shift-invariant spaces,'' \emph{SIAM review}, vol.~43, no.~4, pp. 585--620, 2001. [Online]. Available: \url{https://doi.org/10.1137/S0036144501386986}
\BIBentrySTDinterwordspacing

\bibitem{baron2009bayesian}
\BIBentryALTinterwordspacing
D.~Baron, S.~Sarvotham, and R.~G. Baraniuk, ``Bayesian compressive sensing via belief propagation,'' \emph{IEEE Transactions on Signal Processing}, vol.~58, no.~1, pp. 269--280, 2009. [Online]. Available: \url{https://doi.org/10.1109/TSP.2009.2027773}
\BIBentrySTDinterwordspacing

\bibitem{chen2019quantum}
\BIBentryALTinterwordspacing
Q.-M. Chen, F.~Deppe, R.-B. Wu, L.~Sun, Y.-x. Liu, Y.~Nojiri, S.~Pogorzalek, M.~Renger, M.~Partanen, K.~G. Fedorov \emph{et~al.}, ``Quantum fourier transform in oscillating modes,'' \emph{arXiv preprint arXiv:1912.09861}, 2019. [Online]. Available: \url{https://arxiv.org/abs/1912.09861}
\BIBentrySTDinterwordspacing

\bibitem{PhysRevResearch.2.013012}
\BIBentryALTinterwordspacing
A.~Martin, L.~Lamata, E.~Solano, and M.~Sanz, ``Digital-analog quantum algorithm for the quantum fourier transform,'' \emph{Phys. Rev. Res.}, vol.~2, p. 013012, Jan 2020. [Online]. Available: \url{https://link.aps.org/doi/10.1103/PhysRevResearch.2.013012}
\BIBentrySTDinterwordspacing

\bibitem{boykin2000new}
\BIBentryALTinterwordspacing
P.~Boykin, T.~Mor, M.~Pulver, V.~Roychowdhury, and F.~Vatan, ``A new universal and fault-tolerant quantum basis,'' \emph{Information Processing Letters}, vol.~75, no.~3, pp. 101--107, 2000. [Online]. Available: \url{https://www.sciencedirect.com/science/article/pii/S0020019000000843}
\BIBentrySTDinterwordspacing

\bibitem{dawson2005solovay}
\BIBentryALTinterwordspacing
C.~M. Dawson and M.~A. Nielsen, ``The solovay-kitaev algorithm,'' 2005. [Online]. Available: \url{https://arxiv.org/abs/quant-ph/0505030}
\BIBentrySTDinterwordspacing

\bibitem{qudit2013mischuck}
\BIBentryALTinterwordspacing
B.~Mischuck and K.~M\o{}lmer, ``Qudit quantum computation in the jaynes-cummings model,'' \emph{Phys. Rev. A}, vol.~87, p. 022341, Feb 2013. [Online]. Available: \url{https://link.aps.org/doi/10.1103/PhysRevA.87.022341}
\BIBentrySTDinterwordspacing

\bibitem{Kumar_2025}
\BIBentryALTinterwordspacing
S.~Kumar, N.~N. Hegade, A.-M. Visuri, B.~A. Bhargava, J.~F.~R. Hernandez, E.~Solano, F.~Albarrán-Arriagada, and G.~A. Barrios, ``Digital-analog quantum computing of fermion-boson models in superconducting circuits,'' \emph{npj Quantum Information}, vol.~11, no.~1, Mar. 2025. [Online]. Available: \url{http://dx.doi.org/10.1038/s41534-025-01001-4}
\BIBentrySTDinterwordspacing

\bibitem{sawaya2020resource}
\BIBentryALTinterwordspacing
N.~P. Sawaya, T.~Menke, T.~H. Kyaw, S.~Johri, A.~Aspuru-Guzik, and G.~G. Guerreschi, ``Resource-efficient digital quantum simulation of d-level systems for photonic, vibrational, and spin-s hamiltonians,'' \emph{npj Quantum Information}, vol.~6, no.~1, p.~49, 2020. [Online]. Available: \url{https://doi.org/10.1038/s41534-020-0278-0}
\BIBentrySTDinterwordspacing

\bibitem{dalzell2023quantum}
\BIBentryALTinterwordspacing
A.~M. Dalzell, S.~McArdle, M.~Berta, P.~Bienias, C.-F. Chen, A.~Gily{\'e}n, C.~T. Hann, M.~J. Kastoryano, E.~T. Khabiboulline, A.~Kubica \emph{et~al.}, ``Quantum algorithms: A survey of applications and end-to-end complexities,'' \emph{arXiv preprint arXiv:2310.03011}, 2023. [Online]. Available: \url{https://arxiv.org/abs/2310.03011}
\BIBentrySTDinterwordspacing

\bibitem{Fraser_1965}
\BIBentryALTinterwordspacing
W.~Fraser, ``A survey of methods of computing minimax and near-minimax polynomial approximations for functions of a single independent variable,'' \emph{J. ACM}, vol.~12, no.~3, p. 295–314, jul 1965. [Online]. Available: \url{https://doi.org/10.1145/321281.321282}
\BIBentrySTDinterwordspacing

\bibitem{HomeCharacteristicFunction}
\BIBentryALTinterwordspacing
C.~Fl\"uhmann and J.~P. Home, ``Direct characteristic-function tomography of quantum states of the trapped-ion motional oscillator,'' \emph{Phys. Rev. Lett.}, vol. 125, p. 043602, Jul 2020. [Online]. Available: \url{https://link.aps.org/doi/10.1103/PhysRevLett.125.043602}
\BIBentrySTDinterwordspacing

\bibitem{low2017hamiltonian}
\BIBentryALTinterwordspacing
G.~H. Low and I.~L. Chuang, ``Hamiltonian simulation by uniform spectral amplification,'' 2017. [Online]. Available: \url{https://arxiv.org/abs/1707.05391}
\BIBentrySTDinterwordspacing

\bibitem{Biskit}
\BIBentryALTinterwordspacing
T.~J. Stavenger, E.~Crane, K.~C. Smith, C.~T. Kang, S.~M. Girvin, and N.~Wiebe, ``{C2QA} - {B}osonic {Q}iskit,'' in \emph{2022 IEEE High Performance Extreme Computing Conference (HPEC)}, 2022, pp. 1--8. [Online]. Available: \url{https://ieeexplore.ieee.org/document/9926318}
\BIBentrySTDinterwordspacing

\bibitem{johansson2012qutip}
\BIBentryALTinterwordspacing
J.~Johansson, P.~Nation, and F.~Nori, ``{QuTiP}: An open-source python framework for the dynamics of open quantum systems,'' \emph{Computer Physics Communications}, vol. 183, no.~8, p. 1760–1772, Aug. 2012. [Online]. Available: \url{http://dx.doi.org/10.1016/j.cpc.2012.02.021}
\BIBentrySTDinterwordspacing

\bibitem{hastrup2022universal}
\BIBentryALTinterwordspacing
J.~Hastrup, K.~Park, J.~B. Brask, R.~Filip, and U.~L. Andersen, ``Universal unitary transfer of continuous-variable quantum states into a few qubits,'' \emph{Phys. Rev. Lett.}, vol. 128, p. 110503, Mar 2022. [Online]. Available: \url{https://link.aps.org/doi/10.1103/PhysRevLett.128.110503}
\BIBentrySTDinterwordspacing

\bibitem{bruzewicz2019trapped}
\BIBentryALTinterwordspacing
C.~D. Bruzewicz, J.~Chiaverini, R.~McConnell, and J.~M. Sage, ``{Trapped-ion quantum computing: Progress and challenges},'' \emph{Applied Physics Reviews}, vol.~6, no.~2, p. 021314, 05 2019. [Online]. Available: \url{https://doi.org/10.1063/1.5088164}
\BIBentrySTDinterwordspacing

\bibitem{BlaiscQEDReviewRMP2020}
\BIBentryALTinterwordspacing
A.~Blais, A.~L. Grimsmo, S.~M. Girvin, and A.~Wallraff, ``Circuit quantum electrodynamics,'' \emph{Rev. Mod. Phys.}, vol.~93, p. 025005, May 2021. [Online]. Available: \url{https://link.aps.org/doi/10.1103/RevModPhys.93.025005}
\BIBentrySTDinterwordspacing

\bibitem{PRXQuantum.4.030336}
\BIBentryALTinterwordspacing
O.~Milul, B.~Guttel, U.~Goldblatt, S.~Hazanov, L.~M. Joshi, D.~Chausovsky, N.~Kahn, E.~\ifmmode~\mbox{\c{C}}\else \c{C}\fi{}ifty\"urek, F.~Lafont, and S.~Rosenblum, ``Superconducting cavity qubit with tens of milliseconds single-photon coherence time,'' \emph{PRX Quantum}, vol.~4, p. 030336, Sep 2023. [Online]. Available: \url{https://link.aps.org/doi/10.1103/PRXQuantum.4.030336}
\BIBentrySTDinterwordspacing

\bibitem{simon2020measuring}
\BIBentryALTinterwordspacing
G.~G.~K. Simon, ``Measuring trapped-ion motional decoherence through direct manipulation of motional coherent states,'' Ph.D. dissertation, Massachusetts Institute of Technology, 2020. [Online]. Available: \url{https://hdl.handle.net/1721.1/130221}
\BIBentrySTDinterwordspacing

\bibitem{wang2021single}
\BIBentryALTinterwordspacing
P.~Wang, C.-Y. Luan, M.~Qiao, M.~Um, J.~Zhang, Y.~Wang, X.~Yuan, M.~Gu, J.~Zhang, and K.~Kim, ``Single ion qubit with estimated coherence time exceeding one hour,'' \emph{Nature communications}, vol.~12, no.~1, p. 233, 2021. [Online]. Available: \url{https://doi.org/10.1038/s41467-020-20330-w}
\BIBentrySTDinterwordspacing

\bibitem{devitt2013quantum}
\BIBentryALTinterwordspacing
S.~J. Devitt, W.~J. Munro, and K.~Nemoto, ``Quantum error correction for beginners,'' \emph{Reports on Progress in Physics}, vol.~76, no.~7, p. 076001, 2013. [Online]. Available: \url{https://doi.org/10.1088/0034-4885/76/7/076001}
\BIBentrySTDinterwordspacing

\bibitem{noh2020encoding}
\BIBentryALTinterwordspacing
K.~Noh, S.~Girvin, and L.~Jiang, ``Encoding an oscillator into many oscillators,'' \emph{Physical Review Letters}, vol. 125, no.~8, p. 080503, 2020. [Online]. Available: \url{https://doi.org/10.1103/PhysRevLett.125.080503}
\BIBentrySTDinterwordspacing

\bibitem{christensen2003introduction}
O.~Christensen \emph{et~al.}, \emph{An introduction to frames and Riesz bases}.\hskip 1em plus 0.5em minus 0.4em\relax Springer, 2003, vol.~7.

\bibitem{cleve2000fast}
\BIBentryALTinterwordspacing
R.~Cleve and J.~Watrous, ``Fast parallel circuits for the quantum fourier transform,'' in \emph{Proceedings 41st Annual Symposium on Foundations of Computer Science}.\hskip 1em plus 0.5em minus 0.4em\relax IEEE, 2000, pp. 526--536. [Online]. Available: \url{https://doi.org/10.1109/SFCS.2000.892140}
\BIBentrySTDinterwordspacing

\bibitem{weimann2016implementation}
\BIBentryALTinterwordspacing
S.~Weimann, A.~Perez-Leija, M.~Lebugle, R.~Keil, M.~Tichy, M.~Gr{\"a}fe, R.~Heilmann, S.~Nolte, H.~Moya-Cessa, G.~Weihs \emph{et~al.}, ``Implementation of quantum and classical discrete fractional fourier transforms,'' \emph{Nature Communications}, vol.~7, no.~1, p. 11027, 2016. [Online]. Available: \url{http://dx.doi.org/10.1038/ncomms11027}
\BIBentrySTDinterwordspacing

\bibitem{assouly2023quantum}
\BIBentryALTinterwordspacing
R.~Assouly, R.~Dassonneville, T.~Peronnin, A.~Bienfait, and B.~Huard, ``Quantum advantage in microwave quantum radar,'' \emph{Nature Physics}, vol.~19, no.~10, pp. 1418--1422, 2023. [Online]. Available: \url{https://doi.org/10.1038/s41567-023-02113-4}
\BIBentrySTDinterwordspacing

\bibitem{lanzagorta2011quantum}
M.~Lanzagorta, \emph{Quantum radar}.\hskip 1em plus 0.5em minus 0.4em\relax Morgan \& Claypool Publishers, 2011.

\bibitem{heath2016overview}
\BIBentryALTinterwordspacing
R.~W. Heath, N.~Gonzalez-Prelcic, S.~Rangan, W.~Roh, and A.~M. Sayeed, ``An overview of signal processing techniques for millimeter wave {MIMO} systems,'' \emph{IEEE journal of selected topics in signal processing}, vol.~10, no.~3, pp. 436--453, 2016. [Online]. Available: \url{https://doi.org/10.1109/JSTSP.2016.2523924}
\BIBentrySTDinterwordspacing

\bibitem{gurses2024free}
\BIBentryALTinterwordspacing
V.~Gurses, S.~I. Davis, N.~Sinclair, M.~Spiropulu, and A.~Hajimiri, ``Free-space quantum information platform on a chip,'' \emph{arXiv preprint arXiv:2406.09158}, 2024. [Online]. Available: \url{https://arxiv.org/abs/2406.09158}
\BIBentrySTDinterwordspacing

\bibitem{PhysRevResearch.4.023006}
\BIBentryALTinterwordspacing
S.~Gao, F.~Hayes, S.~Croke, C.~Messenger, and J.~Veitch, ``Quantum algorithm for gravitational-wave matched filtering,'' \emph{Phys. Rev. Res.}, vol.~4, p. 023006, Apr 2022. [Online]. Available: \url{https://link.aps.org/doi/10.1103/PhysRevResearch.4.023006}
\BIBentrySTDinterwordspacing

\bibitem{9951210}
\BIBentryALTinterwordspacing
K.~Miyamoto, G.~Morrás, T.~S. Yamamoto, S.~Kuroyanagi, and S.~Nesseris, ``Gravitational wave matched filtering by quantum monte carlo integration and quantum amplitude amplification,'' in \emph{2022 IEEE International Conference on Quantum Computing and Engineering (QCE)}, 2022, pp. 788--790. [Online]. Available: \url{https://doi.ieeecomputersociety.org/10.1109/QCE53715.2022.00119}
\BIBentrySTDinterwordspacing

\bibitem{10890646}
\BIBentryALTinterwordspacing
S.~R. Nair, B.~Southwell, and C.~Ferrie, ``Short-time quantum fourier transform processing,'' in \emph{ICASSP 2025 - 2025 IEEE International Conference on Acoustics, Speech and Signal Processing (ICASSP)}, 2025, pp. 1--5. [Online]. Available: \url{https://doi.org/10.1109/ICASSP49660.2025.10890646}
\BIBentrySTDinterwordspacing

\end{thebibliography}


\end{document}